\documentclass[aps,pre,reprint,superscriptaddress,showpacs,preprintnumbers,amsmath,amssymb,nofootinbib]{revtex4-2}
\usepackage[dvipdfmx]{graphicx}
\usepackage{dcolumn}   
\usepackage{bm}        
\usepackage{amssymb, amsmath, amsfonts}   
\usepackage{lipsum}        
\usepackage{color}
\usepackage{appendix}
\usepackage[normalem]{ulem}
\DeclareGraphicsExtensions{{.pdf}}
\begin{document}
\newcommand{\noter}[1]{{\color{red}{#1}}}
\newcommand{\noteb}[1]{{\color{blue}{#1}}}
\newcommand{\noteg}[1]{{\color{green}{#1}}}
\newcommand{\field}{\left( \boldsymbol{r}\right)}
\newcommand{\paren}[1]{\left({#1}\right)}
\newcommand{\vect}[1]{\boldsymbol{#1}}
\newcommand{\uvect}[1]{\tilde{\boldsymbol{#1}}}
\newcommand{\vdot}[1]{\dot{\boldsymbol{#1}}}
\newcommand{\vder}{\boldsymbol{\nabla}}
\widetext
\title{
What Do Deep Neural Networks Find in Disordered Structures of Glasses?
}
\author{Norihiro Oyama}
\email{Norihiro.Oyama.vb@mosk.tytlabs.co.jp}
\affiliation{Toyota Central R\&D Labs, Inc., Bunkyo-ku, Tokyo 112-0004, Japan}

\author{Shihori Koyama}
\affiliation{Toyota Central R\&D Labs, Inc., Bunkyo-ku, Tokyo 112-0004, Japan}

\author{Takeshi Kawasaki}
\email{kawasaki@r.phys.nagoya-u.ac.jp}
\affiliation{Department of Physics, Nagoya University, Nagoya 464-8602, Japan}

\date{\today}
\begin{abstract}
Glass transitions are widely observed in various types of soft matter systems. 
However, the physical mechanism of these transitions remains elusive despite years of ambitious research. In particular, an important unanswered question is whether the glass transition is accompanied by a divergence of the correlation lengths of the characteristic static structures. In this study, we develop a deep-neural-network-based method that is used to extract the characteristic local meso-structures solely from instantaneous particle configurations without any {information} about the dynamics. We first train a neural network to classify configurations of liquids and glasses correctly. Then, we obtain the characteristic structures by quantifying the grounds for the decisions made by the network using Gradient-weighted Class Activation Mapping (Grad-CAM). We consider two qualitatively different glass-forming binary systems, and through comparisons with several established structural indicators, we demonstrate that our system can be used to identify characteristic structures that depend on the details of the systems. Moreover, the extracted structures are remarkably correlated with the nonequilibrium aging dynamics in thermal fluctuations.
\end{abstract}
\maketitle
%
\section{Introduction}
When a liquid is cooled while preventing crystallization by quenching or adding impurities, a liquid state can be maintained below melting temperature, resulting in a so-called supercooled liquid state. 
Further cooling of the supercooled liquid results in a dramatic increase in the viscosity of the liquid and yields a glass (more generally, an amorphous solid). In such a system, the particle motion is frozen, and the structure remains disordered. 
Various materials, e.g. oxides, alloys, polymers, and colloids, take on glassy states. 
Glassy materials are \textit{generally} considered disordered and homogeneous because they {basically} cannot be distinguished from simple liquids that are also disordered in structure using analytical methods such as neutron, X-ray, or light scattering and other two-body correlations in the density field. However, dramatic changes to their properties can occur, for example, a 15-order-of-magnitude increase in the viscosity from a temperature change of only approximately 20\%{~\cite{Angell1995}}. 
Although the glass transition phenomenon has been studied for more than 150 years, its mechanism has not yet been clarified~\cite{Debenedetti2001,Dyre2006,Cavagna2009,Berthier2011}.

Heterogeneity in particle motion develops in supercooled liquids near the glass transition temperature, and the spatial length scale increases on such a glass transition~\cite{Kob1997,Yamamoto1998,Yamamoto1998a,Lacevic2003,Dauchot2005,Karmakar2009,Keys2011}. 
This behavior is called dynamic heterogeneity and is a potential cause for the rapid increase in viscosity at the glass transition point. However, to date, the origin of this dynamic heterogeneity has not been clarified; in particular, questions remain as to whether it is formed entirely dynamically or whether a static structure exists in the background. 
The ``dynamical facilitation theory'' describes the heterogeneity associated with glass transitions as a fully dynamic phenomenon, and explains the experimental results and numerical analysis of glass transitions~\cite{Chandler2010}. In contrast, ``the theory of random first-order transition'' (RFOT), which considers the glass transition as a thermodynamic phase transition and proposes a scenario in which a static conceptual structure called a ``mosaic'' develops, also explains the experimental results and numerical analysis of glass transitions~\cite{Kirkpatrick1989,Cavagna2009}. 
{Thus, although these theories are contradictory in terms of whether the glass transition is a purely dynamic transition or a thermodynamic phase transition governed by a static structure, they can explain various aspects of the glass transition phenomenon. 
Hence, in the current state, there appears to be no definitive theory for understanding the full picture of glass transitions.}

Many attempts have been made to explore the specific structures that exist in supercooled liquids. For instance, icosahedral-like structures in metallic glasses~\cite{Hirata2011,Hirata2013} and medium-range crystalline order in colloidal glasses with small particle-size dispersity have been found~\cite{Kawasaki2007,Kawasaki2010,Kawasaki2011,Leocmach2012}. Order parameters are introduced on a system-by-system basis to extract these characteristic structures, but no order parameter applicable to \textit{all} amorphous solids has been found. It is also unclear whether such characteristic structures are universal; this is a topic of active debate. Therefore, elucidating the presence or absence of a universal structure in amorphous solids is a significant and challenging problem in fundamental physics. 
Tong and Tanaka recently developed a new order parameter consisting of the bond angles of particle structures and successfully extracted the characteristic structures correlated with particle dynamics for a wide range of glass-forming systems, including binary mixtures and polydisperse glassy systems with large particle size dispersions{~\cite{TTPRX,TTNatCommun}}. However, as indicated in the corresponding literature~\cite{TTPRX,TTNatCommun}, this method has not been able to extract the characteristic structures in the Kob--Andersen system~\cite{KAM}, a typical glass-forming model, and consequently, attempts to develop a universal structural analysis method for a variety of glass-like systems continues to the present day.
{We mention that as another branch of examples of promising static information-based approaches, a method relying on the Franz-Parisi potential has been proposed~\cite{Berthier2021PRL}. 
Its effectiveness was demonstrated by the quantitative correspondence with the structural relaxation time.
The recently proposed microscopic version of a similar Franz-Parisi potential-based quantity would allow us to specify the local characteristic structures that govern the dynamics on the purely static basis~\cite{FP_JCP2022}.}

In recent years, machine-learning approaches have been widely used to investigate the characteristic structures governing glass dynamics~\cite{Schoenholz2017,Boattini2020,Bapst2020,Hayato,Boattini2020,Paret2020}. 
In particular, recent studies have successfully predicted the dynamics from the static structure in Kob--Andersen systems by learning from a large amount of structural data, as well as the corresponding dynamic data, using graph neural networks~\cite{Bapst2020,Hayato}. 
In addition to these supervised approaches, unsupervised counterparts have also been applied to the extraction of characteristic structures from glasses, pioneered by Ronhovde and co-workers~\cite{Ronhovde2011,Ronhovde2012}. 
Interestingly, many researchers have recently reported that the structures extracted using unsupervised methods~\cite{Boattini2020,Paret2020} exhibit correlations with the long-time dynamics, despite no information about the dynamics being provided during the training.
However, although machine learning is very promising for exploring the structures of glasses, accurate learning (including preparation of the training data) is computationally expensive, and the results are difficult to interpret.

In this work, we propose a method to extract the characteristic multi-particle structures of glasses solely from the static configurations using a deep learning-based approach. To this end, we work on the classification problem for the random structures in glasses and liquids using a convolutional neural network (CNN)~\cite{CNN} and then identify the structures that the CNN relied on to make decisions using gradient-weighted class activation mapping (Grad-CAM)~\cite{GradCAM}. We applied our proposed method to two representative glass-forming liquid systems and compared the obtained structures with well-established structural indicators. The results demonstrate that the proposed method can extract qualitatively different characteristic structures in a system-detail-dependent manner. 
{Surprisingly}, although our method does not refer to information about the dynamics during the learning process and extracts the characteristic structure solely from the instantaneous static configurations, the obtained structures strongly correlate with the nonequilibrium aging dynamics.

The remainder of this paper is organized as follows. 
In Sec.~\ref{sec:simulations}, we summarize the simulation methods and protocols used for sample preparation for the deep-learning tasks. 
In Sec.~\ref{sec:analyses}, we introduce the CNN and Grad-CAM, and provide a brief explanation of the established structural indicators used as a reference. 
In Sec.~\ref{sec:results}, the results of the structural analyses are presented, and the correlation between distinct indicators, as well as the predictability of our method with respect to the dynamics, is discussed. 
Finally, in Sec.~\ref{sec:summary}, we provide a summary of this study and an overview of future research directions.

\section{Simulations}\label{sec:simulations}
\subsection{System setups}
In this study, we consider two distinct systems: the Kob--Andersen model (KAM)~\cite{KAM} and the additive binary mixture (ABM)~\cite{Oyama2021a}. 
Both systems are two-dimensional (2D) and are described by the Lennard--Jones (LJ) potential with linear smoothing terms:
\begin{align}
    \phi_{ij}(r)&=\phi_{ij}^\ast(r)-\phi_{ij}^\ast(r_{ij}^c)-(r-r_{ij}^c)\frac{d\phi_{ij}^\ast}{dr}|_{r=r_{ij}^c},\label{eq:potential}\\
    \phi_{ij}^{\ast}(r)&=4\epsilon_{ij}\left[\left(\frac{\sigma_{ij}}{r}\right)^{12}-\left(\frac{\sigma_{ij}}{r}\right)^{6}\right],
\end{align}
where the subscript $ij$ indicates that the variable is between particles $i$ and $j$, $\epsilon_{ij}$ sets the energy scale, $\sigma_{ij}$ determines the interaction range, and {$r_{ij}^c$} is the cutoff length. The dynamics of the particles obey $\phi_{ij}$, whereas $\phi_{ij}^\ast$ is the reference standard LJ potential. The smoothed LJ potential $\phi$ guarantees the continuity of the potential and force at cutoff distance $r=r_{ij}^c$, thus eliminating undesired artifact effects owing to the introduction of the cutoff~\cite{Shimada}.

Both systems are composed of two different types of particles (A and B) and are characterized by different parameter sets, such as $\epsilon_{ij}$ and $\sigma_{ij}$. 
In the case of the KAM, the LJ parameters are nonadditive: $\sigma_{AA}=1$, $\sigma_{AB}=0.8$, $\sigma_{BB}=0.88$, $\epsilon_{AA}=1$, $\epsilon_{AB}=1.5$, and $\epsilon_{BB}=0.5$. Therefore, the concept of ``particle size'' is not well defined in the KAM system.
For the ABM, on the other hand, the parameters are simply additive; thus, we can unambiguously say that particle A is small, and B is large (that is, $\sigma_{AA}=5/6, \sigma_{AB}=1, \sigma_{BB}=7/6$, and $\epsilon_{ij}=1$) regardless of the combination of types of particles $i$ and $j$. 

All observables were nondimensionalized using characteristic length $\sigma^\ast$, characteristic energy $\epsilon^\ast$, and particle mass $m^\ast$ (the characteristic variables are listed in Table~\ref{table:parameters}). 
{The total number of particles was fixed at $N=N_A+N_B=2000$.
The number density $\rho=N/L^2$ and the number ratio of the two-particle species $N_A/N_B$ also differ between the two systems, i.e., the KAM and ABM. }
With the values of $\rho$ used here, the systems entered the glassy phase once the temperature was sufficiently low~\cite{KAM,Oyama2021a}.
Information about the parameters mentioned here is summarized in Table~\ref{table:parameters}.
{Although we consider only 2D systems in this article for the sake of simplicity, we stress that all the analyses here can be easily extended to three-dimensional systems, which will be performed in the future.}
\tabcolsep = 5pt
\begingroup
\renewcommand{\arraystretch}{1.5}
\begin{table*}[htb]
  \caption{Summary of model parameters}
  \centering\label{table:parameters}
  \begin{tabular}{c|ccccccccccccccc}
    \hline
      & $m_{A}$& $m_{B}$& $\sigma_{AA}$& $\sigma_{AB}$& $\sigma_{BB}$& $\epsilon_{AA}$& $\epsilon_{AB}$& $\epsilon_{BB}$ & $N_A$& $N_B$ &$\rho$ &$\sigma^\ast$ &$\epsilon^\ast$ &$m^\ast$ &$r_{ij}^c$\\
    \hline 
    KAM &$1.0$ &$1.0$ &$1.0$ &$0.8$ &$0.88$ &$1.0$ &$1.5$ &$0.5$ &$1300$ &$700$ &$1.2$ &$\sigma_{AA}$ &$\epsilon_{AA}$ &$m_{A}$ &$2.5\sigma_{ij}$\\
    ABM &$1.0$ &$1.0$ &$5/6$ &$1.0$ &$7/6$ &$1.0$ &$1.0$ &$1.0$ &$1000$ &$1000$ &$1.09$ &$\sigma_{AB}$ &$\epsilon_{AB}$ &$m_{A}$ &$3.0\sigma_{ij}$\\
\hline
  \end{tabular}
\end{table*}
\endgroup

\subsection{Sample preparation protocol}
We performed molecular dynamics (MD) simulations using the open-source Large-scale Atomic/Molecular Massively Parallel Simulator (LAMMPS: {https://www.lammps.org/}). We generated samples via NVT simulations using the Nos\'{e}--Hoover thermostat. Periodic boundary conditions were set in all directions. In this study, we aim to address a simple binary classification problem. For the two classes, we chose configurations at the temperature where the dynamic slowing-down starts at ($T_{\rm L}=0.8$ for KAM and $T_{\rm L}=2.0$ for ABM) and at a very low temperature ($T_{\rm G}=0.05$ for both models). 
For both KAM and ABM, we first generated 5000 independent random configurations (4000 were used for training, 400 for validation, and 600 for the test data) and equilibrated them at a very high temperature of $T=4.0$. The obtained configurations were then cooled at a constant cooling rate ($\dot{T}\approx 8.33\times 10^{-5}$), and the samples for the classification tasks were obtained at the desired temperatures $T_{\rm L}$ and $T_{\rm G}$. The samples at $T_{\rm L}$ correspond to ``equilibrium'' supercooled liquids in the sense that their dynamics exhibit time-translational invariance, whereas those at $T_{\rm G}$ are regarded as ``nonequilibrium'' glasses in the sense that they are expected to experience aging. Note that, judging from the evolution of the potential energy of the system as a function of the temperature (Fig.~S6 in the Supplemental Material)~\cite{Sastry1998}, crystallization is avoided in both systems at this cooling rate (i.e., we did not observe any discontinuous jumps in the energy). 
Consequently, the radial distribution function $g(r)$ of the configurations at $T_{\rm G}$ does not show any signs of global crystallization (Fig.~S7 in the Supplemental Material).

\section{Analytical methods}\label{sec:analyses}
{We train a neural network to distinguish two classes of systems, ``glass'' and ``liquid,'' based purely on the instantaneous configurations.
Then, the characteristic structures of glasses are identified by extracting the meso-scale structures that the trained network relied on to provide a correct classification.
In this section, we explain the methods used to achieve these classifications and identifications of characteristic structures.}

\subsection{Convolutional Neural Network (CNN)}\label{sec:CNN}
{We first perform supervised learning to train a CNN~\cite{CNN,Swanson2020} to predict whether a given configuration is glass or liquid. }
Following ref.~\cite{Swanson2020} in which the authors tackled a similar classification task successfully, our network has no pooling layers. 
It is then simply composed of three convolutional layers and subsequent activations (the rectified linear units), followed by the fully connected layer, dropout layer, and final fully connected layer as the output layer. 
Note that, although we apply the ``softmax'' function subsequently to obtain the final results, the output layer of the network is the fully connected one to make it compatible with Grad-CAM, as explained in Sec.~\ref{subsec:gradCAM}. 
The full details of the network and learning protocol, including the precise values of the hyperparameters, are summarized in the Supplemental Material.
{After training, the softmax layer outputs a value in the range of [0,1] which can then be interpreted as the probability for a configuration to be assigned to one of these classes.}

Importantly, when we feed the particle configurations, $\rho(\boldsymbol{r})=\sum_i^N\delta(\boldsymbol{r}-\boldsymbol{r}_i)$,
obtained from the MD simulations into the CNN, they are gridized by the mapping operator ${\cal M}$: $\tilde{\rho}\equiv {\cal M}(\rho)$ (see Supplemental Material for technical details). Here, the tilde indicates that the variable is grid-based. We also mention that in the padding process at the convolutional layers, we use circular padding to properly consider the periodic boundary conditions.

\subsection{Gradient-weighted class activation mapping (Grad-CAM)}\label{subsec:gradCAM}
{Once a CNN is able to classify glasses and liquids correctly after training, we aim to extract the characteristic mesoscale structures that the CNN relies on when classifying.}
This identification of crucial information is called ``class activation mapping'' (CAM).
The first-proposed simple CAM~\cite{CAM} assumed a global average pooling at the end of the network, and thus, cannot be utilized for networks with different types of architectures. This problem has been solved using a method called gradient-weighted class activation mapping (Grad-CAM)~\cite{GradCAM}. In Grad-CAM, CAM is calculated based on the differential of the output of the network with respect to the feature maps as
\begin{align}
    \alpha_m^C &= \frac{1}{Z}\sum_{k}^u\sum_{l}^v\frac{\partial y^C}{\partial A_{k,l}^m},\\
    \tilde{L}^C &= ReLU(\sum_m\alpha_m^C \tilde{A}^m),
\end{align}
where $y^C$ is the score for class $C$ ($C\in\lbrace {glass},{liquid}\rbrace$ in the current setup) before softmax, $\tilde{A}^m$ is the $m$-th feature map activation of a convolutional layer, $A_{k,l}^m$ is the $(k,l)$ component of $\tilde{A}^m$, and $Z=uv$ is the normalizing factor for the global pooling calculation. The rectified linear unit $ReLU$ simply returns $x$ if $x>0$ and zero otherwise. Thus, in this Grad-CAM method, the characteristic part of the input information is identified as the weighted sum of the feature maps after a specified convolution layer (usually the last layer), and the weights are obtained depending on the global average of the sensitivity (gradient) of the output with respect to each pixel of the feature maps. Importantly, this method can be applied to networks with any architecture if the backpropagation is tractable.

The results presented in this paper are all particle-based Grad-CAM scores, $\Gamma=\sum_i^N \Gamma_i\delta(\boldsymbol{r}-\boldsymbol{r}_i)$, obtained using the inverse mapping operator ${\cal M}^{-1}$: $\Gamma\equiv {\cal M}^{-1}(\tilde{L}^C)$, where $\Gamma_i$ is the Grad-CAM score of particle $i$.
We simply call $\Gamma$ the Grad-CAM score. {Note that, hereinafter, all particle-based variables are coarse-grained and normalized (see the Supplemental Material for technical details, including the precise definition of $\Gamma_i$).}

\subsection{Voronoi volume}
In this study, we compared the results of the proposed method with those of handcrafted structural indicators to interpret the obtained Grad-CAM score $\Gamma$.
The first reference indicator is the volume of the Voronoi cells $\Upsilon$ that particles reside in (here, we call them volumes, although they are in fact areas because the system is 2D).
{The volume of the Voronoi cell allows us to quantify the so-called free volume of each particle, which
is considered a significant static characteristic that explains the divergence of the viscosity in glass transition (the free volume theory)~\cite{Cohen1959}.
We note that, in ref.~\cite{Widmer-Cooper2005,Widmer-Cooper2006}, the {\it microscopic} correlation between the free volume and the dynamics (i.e., the dynamical propensity) was studied and concluded to be not significantly correlated. 
{On the contrary, a strong correlation between the free volume and bond-bond breakage occurring over long periods of time in low-temperature glassy systems has been reported~\cite{Shiba2013}. Despite this controversial situation,} because there is no doubt that the free volume of the particles is an important interpretable static property determined geometrically from the particle structure, we will refer to it here as one of the structural indicators.}

The Voronoi cell to which particle $i$ belongs can be uniquely defined without the introduction of any additional parameters, as follows:
\begin{align}
    V(\boldsymbol{r}_i) = \lbrace \boldsymbol{r}|{\cal D}(\boldsymbol{r},\boldsymbol{r}_i)\le {\cal D}(\boldsymbol{r},\boldsymbol{r}_j), j\ne i\rbrace,
\end{align}
where ${\cal D}(\boldsymbol{a},\boldsymbol{b})$ is a function that provides the 2D Euclidean distance between points $\boldsymbol{a}$ and $\boldsymbol{b}$.
The point $\boldsymbol{r}$ in the equation is an arbitrary point in the system that is independent of the particle density field $\rho(\boldsymbol{r})$. 
The volume of the Voronoi cell for particle $i$ can then be obtained as $\Upsilon_i\equiv{\cal V}(V(\boldsymbol{r}_i))$, where ${\cal V}(V)$ is the operator that outputs the volume of the region $V$.
To achieve Voronoi tessellation, we used the \emph{freud}~\cite{freud1,Dice2019} Python library, which properly considers periodic boundary conditions.
{The Voronoi cell volumes provide a quantitative measure of the (inverse) local packing density.}
We call $\Upsilon=\sum_i^N \Upsilon_i\delta(\boldsymbol{r}-\boldsymbol{r}_i)$ the Voronoi volume.

\subsection{Tong--Tanaka order parameter}\label{sec:ttop}
{Tong and Tanaka~\cite{TTPRX,TTNatCommun} {proposed an excellent order parameter that can characterize structures correlated with long-time dynamics even in glass-forming systems with large particle size dispersion, where characteristic structures are difficult to capture with bond-orientation order parameters~\cite{Kawasaki2011}}.
We call this order parameter the Tong--Tanaka order parameter (TT-OP). 
The TT-OP has been successfully used as a structural indicator of the dynamic properties of various glasses {and hence we measure it as a reference below.}}

To calculate the TT-OP, we first look at each particle (e.g., particle $o$) and its neighbors (the particles sharing the edges of the Voronoi cell with center particle $o$). Then, particle $o$'s TT-OP, $\Theta_o$, is obtained as the average difference between the angle formed by particle $o$ and two of its neighbors that are adjacent to each other, $\theta^1_{ij}$, and the corresponding ideal angle $\theta^2_{ij}$ (i.e., that is obtained when the distances between particles are exactly the same as the sum of their ``radii''; see the Supplemental Material for more details and a schematic of the definitions of $\theta^1_{ij}$ and $\theta^2_{ij}$):
\begin{align}
    \Theta_o=\frac{1}{N_o}\sum_{\langle ij\rangle}|\theta^1_{ij}-\theta^2_{ij}|,
\end{align}
where $N_{o}$ is the number of particles neighboring particle $o$ (this number agrees with the number of neighboring pairs of neighbors).
The TT-OP $\Theta=\sum_i^N \Theta_i\delta(\boldsymbol{r}-\boldsymbol{r}_i)$ is defined as a particle-based indicator, and {it has been shown}
 that, for various systems, the spatially coarse-grained TT-OP predicts the dynamic propensity {very well}~\cite{TTNatCommun}.
{The results (not only of the TT-OP but also of all particle-based variables, including the Grad-CAM score $\Gamma$) presented below are all spatially coarse-grained (and further normalized to the range $[0,1]$).
We explain the coarse-graining procedure in detail in Sec.~\ref{sec:CG}.
}


{\subsection{Dynamic propensity}\label{sec:propensity}
As a measure of the dynamic heterogeneity that appears originated from a specific configuration of particles, the so-called dynamic propensity is usually employed~\cite{Widmer-Cooper2004,Widmer-Cooper2007,Berthier2007}.
To define this variable, we introduce the isoconfigurational ensemble first:
in this special ensemble, samples share an identical initial particle configuration, $\rho_0=\sum_i^N\delta(\boldsymbol{r}-{\bf r}_i(0))$, but have different realizations of velocities with a specified temperature $T$ (the statistics of velocities obeys the Maxwell-Boltzmann distribution with this temperature).
In this study, for each initial configuration, we performed MD simulations with 30 different initial velocity distributions.
For each realization, we calculated 
{intensity of the so-called cage-relative displacement~(CRD)}
\footnote{{We employed the cage-relative displacements to exclude the undesired effects due to anomalous fluctuations that are specific to two-dimensional systems~\cite{Shiba2016PRL,MW_PNAS,Shiba_2018,Shiba2019PRL}.
}}, $\Delta_i^{s}(t)$, which is defined as $\Delta_i^{s}(t)=\sqrt{(\boldsymbol{d}_i^s(t)-\bar{\boldsymbol{d}}_i^{s}(t))^2}$, where $\boldsymbol{d}^s_i(t)\equiv\boldsymbol{r}^s_i(t)-\boldsymbol{r}_i(0)$ is the displacement vector of the particle $i$ at time $t$, $\bar{\boldsymbol{d}}_i^{s}\equiv\frac{1}{N_i}\sum_j\boldsymbol{d}^s_j$ (the sum for $j$ runs over the neighbors of $i$ including $i$ itself, and $N_i$ stands for the number of particles involved here) is the average displacement vector of the cage to which particle $i$ belongs, and the superscript $s$ is the sample index (which distinguishes different realizations of the velocity distribution at $t=0$). 
The dynamic propensity field $\Delta$ is then defined as the average of the sample-dependent values of the {CRD} field, $\Delta^s=\sum_i^N\Delta^{s}_i\delta(\boldsymbol{r}-\boldsymbol{r}_i(0))$, over $N_s$ samples as $\Delta\equiv \frac{1}{N_s}\sum_s^{N_s}\Delta^{s}$.
{As the value of $N_s$, we basically employed $N_s=30$ unless stated otherwise.} 
 
}

{\subsection{Coarse-graining of particle-base indicators}\label{sec:CG}
In refs.~\cite{TTPRX,TTNatCommun}, Tong and Tanaka showed that, when properly coarse-grained, the TT-OP introduced in Sec.~\ref{sec:ttop} correlates strongly with the dynamic propensity field.
In our analyses, we have coarse-grained all the particle-based indicators by a similar method to the one proposed in ref.~\cite{TTNatCommun}:
{\begin{align}
    \bar{X}_i(\xi_X) = \frac{\sum_j X_jP(|\boldsymbol{r}_j-\boldsymbol{r}_i|)}{\sum_j P(|\boldsymbol{r}_j-\boldsymbol{r}_i|)},\label{eq:CG}
\end{align}}
where 
{$P(r)=\exp(-(r/\xi_X)^2)$}
is the coarse-graining kernel and $\xi_X$ is the coarse-graining length
{for variables $X\in\lbrace\Gamma,\Upsilon,\Theta,\Delta\rbrace$}.
For the calculation of the coarse-graining of the {variables $X$, the cutoff distance $r_X^{\rm c}$ in $P(r)$ is introduced, which is fixed as $r_X^{\rm c}=2\xi_X$} in this study.
We employed this coarse-graining procedure (it is slightly different from the one employed in ref.~\cite{TTNatCommun}) after comparing several options.

We stress that, in this work, we coarse-grain not only the structural indicators but also the dynamic propensity field.
We explain how we determine the coarse-graining lengths $\xi_X$ in Sec.~\ref{sec:xi}.
Additionally, all particle-based variables are further normalized to [0, 1] by simply subtracting the minimum value and then dividing by the maximum.
}

\section{Results \& discussions}\label{sec:results}
\subsection{Extraction of characteristic structures using Grad-CAM}
\tabcolsep = 5pt
\begingroup
\renewcommand{\arraystretch}{1.5}
\begin{table}[tb]
  \caption{Classification accuracy for test data}
  \centering\label{table:accuracy}
  \begin{tabular}{cccc}
 
    \hline
     KAM(G) & KAM(L)  & ABM(G) & ABM(L) \\
    \hline 
    {$1.00$} &$0.998$ &$1.00$ &$1.00$\\
    \hline
  \end{tabular}
\end{table}
\endgroup

The CNN introduced in Sec.~\ref{sec:CNN} was run over 250 epochs.
During the training process, 4,000 training data samples (for both glasses and liquids: 8,000 samples in total) were provided with the correct labels indicating whether the samples were glasses or liquids.
For both systems (KAM and ABM), the learning stage proceeded smoothly, and the classification accuracy reached almost 100\% both for the training and validation data after these relatively small epochs. The same degree of accuracy was achieved for the test data (the results for the test data are summarized in Table.~\ref{table:accuracy}).
We stress here that the calculation cost for the training part is very low in our setup: the entire 250 epochs of learning only took {approximately 8 h} using an NVIDIA Quadro P4000 (GP104GL).

\begin{figure*}[tb]
    \centering
    \includegraphics[width=\linewidth]{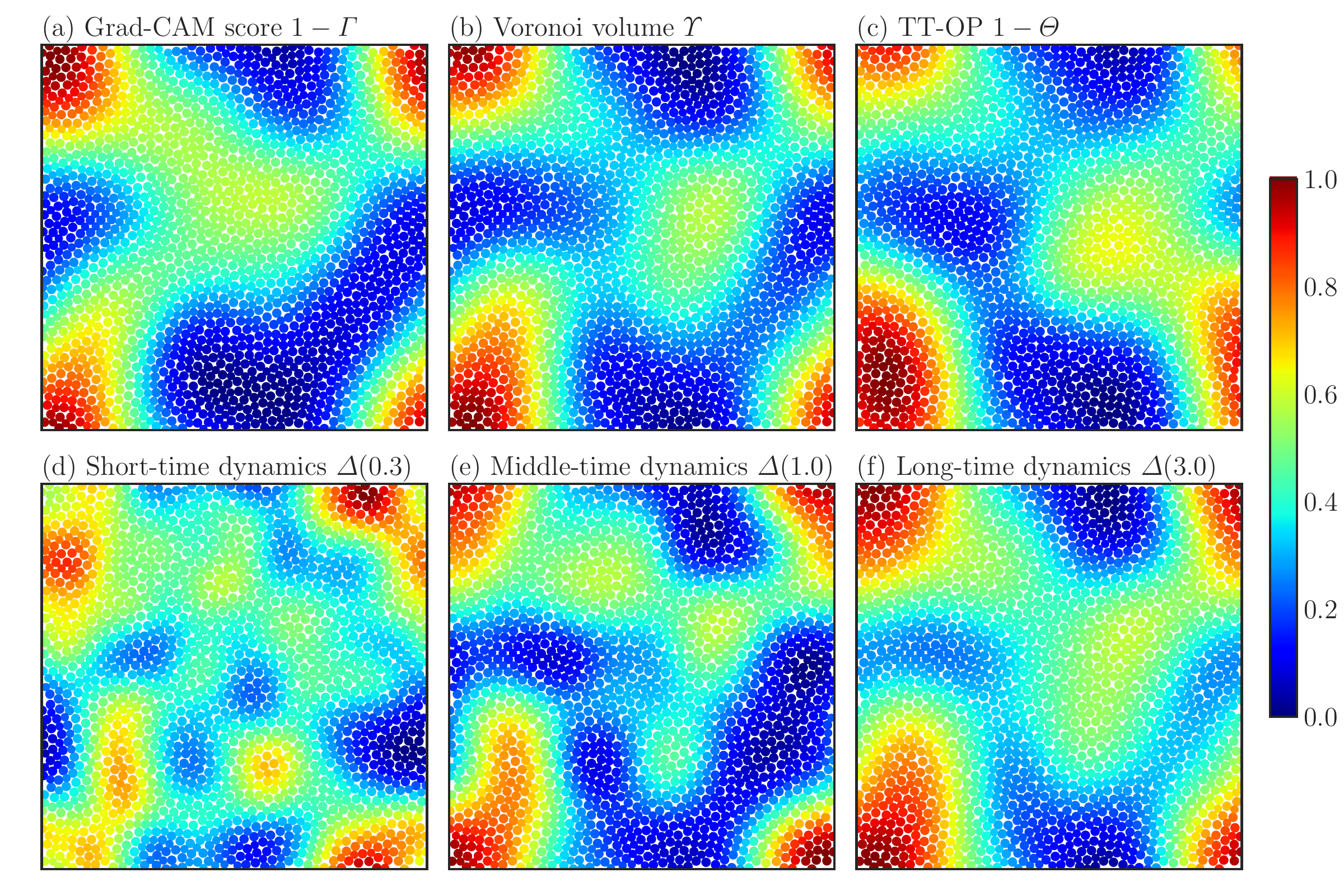}
    \caption{
    Visualization of particle-based structural indicators for a typical glass configuration of the KAM system: (a) Grad-CAM score ($\Gamma$), (b) Voronoi volume ($\Upsilon$), (c) TT-OP ($\Theta$), and (d-f) Dynamic propensity ($\Delta$) {at different ``time'' scales. {The argument of $\Delta$ stands for the mean intensity of the displacements $\delta$ at which $\Delta$ is measured: as indicated in the panel titles, $\delta\approx 0.3,1.0,3.0$,} which roughly corresponds to $t\approx \tau_\alpha, 3\tau_\alpha,10\tau_\alpha$, are employed.}
Notice that all indicators are normalized to [0,1], and the different colors distinguish the values as shown in the color bar. 
    In addition, $1-\Gamma$ and $1-\Theta$ are shown in panels (a) and (c), respectively, rather than $\Gamma$ and $\Theta$, for ease of comparison with the dynamic propensity.
    The precise values of coarse-graining length $\xi_X$ employed here are summarized in Table~\ref{table:xi_values}.
}
    \label{fig:indicators_KAM}
\end{figure*}

\begin{figure*}[tb]
    \centering
    \includegraphics[width=\linewidth]{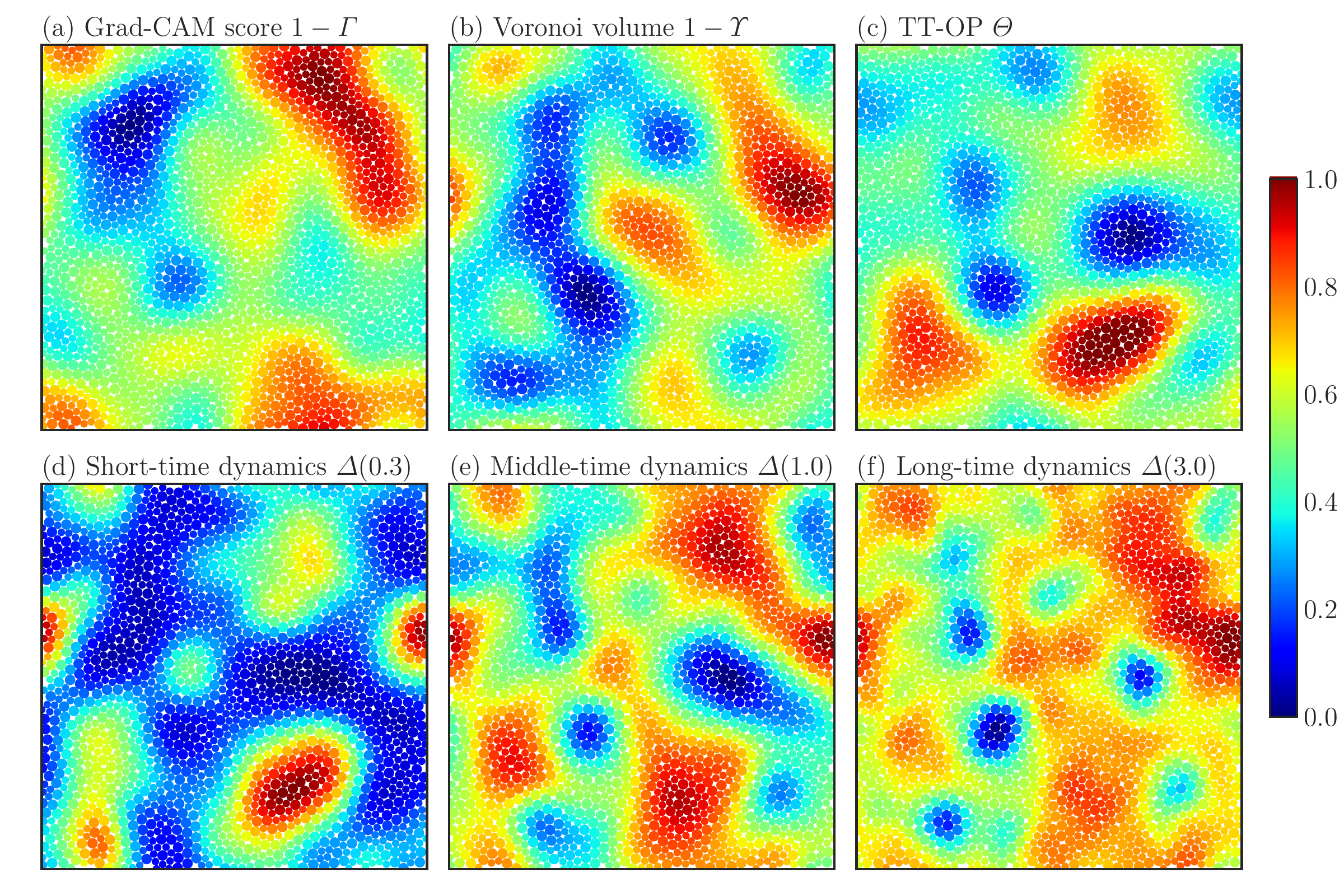}
    \caption{
    Visualization of particle-based structural indicators for a typical glass configuration in the ABM system.
    The meanings of the panels are basically the same as those presented in Fig.~\ref{fig:indicators_KAM}, while $1-\Upsilon$ and $\Theta$ are shown in panels (b) and (c), rather than $\Upsilon$ and $1-\Theta$.}
    \label{fig:indicators_ABM}
\end{figure*}

Subsequently, using Grad-CAM, we further extracted the characteristic structures that the CNN relied on when identifying glass samples as \emph{glasses}.
Notably, this calculation requires only a trivial cost (much less than a second for each sample). 
We present the typical results obtained for the KAM system in Fig.~\ref{fig:indicators_KAM}(a) and those of ABM in Fig.~\ref{fig:indicators_ABM}(a) {(notice that $\Gamma$ visualized here are coarse-grained with the length $\xi_\Gamma$ determined in the next subsection \ref{sec:xi})}. 
Both these results are for the glass configurations: although we can also investigate the characteristic structures of liquids and try to extract glass-like structures from liquids (and \emph{vice versa}) within the Grad-CAM framework, we restricted ourselves to the investigation on the characteristic structures of glasses in this study.
\footnote{Importantly, we discarded a test sample for which the trained CNN gave the wrong classification {(only 1 out of 2,400 samples: in the case of liquid in the KAM)} to rule out the possible influence from such an abnormal sample, e.g., when evaluating the probability distribution function shown later in Fig.~\ref{fig:correlations}. 
However, we would like to mention that the investigation of such samples is still important since they can reside in the vicinity of the ``boundary'' between glass and liquid classes and thus provide meaningful information about their structural differences.
This investigation should be performed as future work.
}


\begin{figure}[tb]
    \centering
    \includegraphics[width=\linewidth]{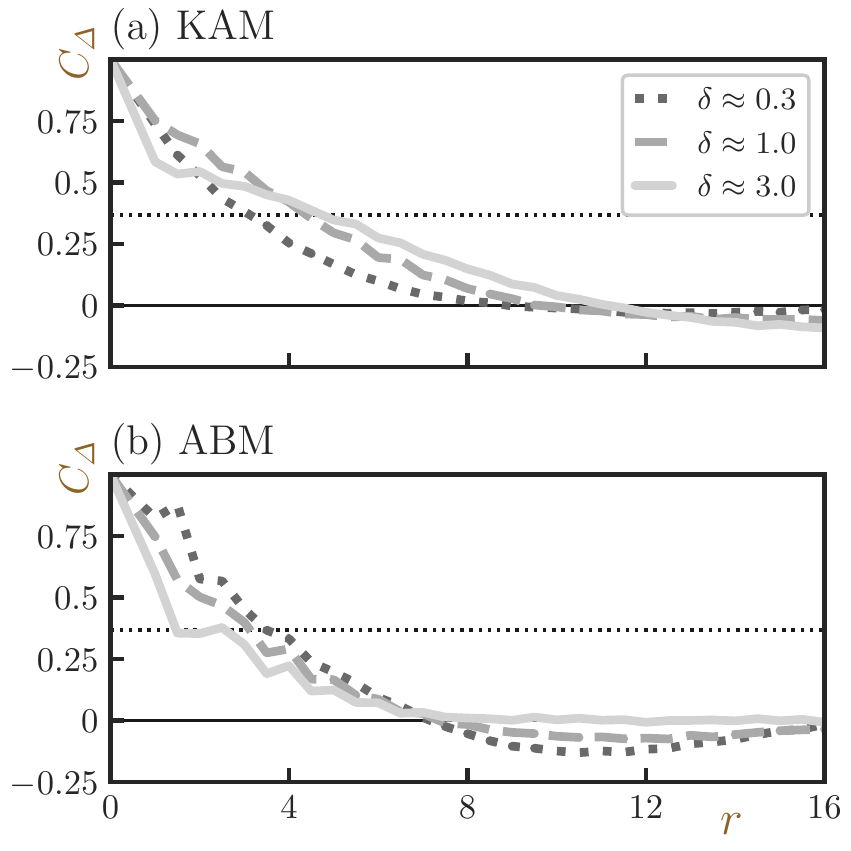}
    \caption{
{The spatial correlation function of the dynamic propensity $\Delta$, $C_\Delta$, as a function of the distance from the reference particle $r$.
Results for (a) KAM and (b) ABM systems.
Different line styles distinguish different time scales as shown in the legend.
The horizontal dotted lines depict $C_\Delta=1/e$.}
}
    \label{fig:dynamics_spatial_corre}
\end{figure}

\tabcolsep = 5pt
\begingroup
\renewcommand{\arraystretch}{1.5}
\begin{table*}[tb]
  \caption{Coarse-graining length for each variable}
  \centering\label{table:xi_values}
  \begin{tabular}{c|ccc|c|ccc|ccc}
  \hline
    &\multicolumn{3}{c|}{Figs.~\ref{fig:indicators_KAM}, \ref{fig:indicators_ABM}}    &\multicolumn{1}{c|}{Figs.~\ref{fig:dynamics},\ref{fig:dynamics_various}}
    &\multicolumn{3}{c|}{Figs.~\ref{fig:indicators_KAM}, \ref{fig:indicators_ABM},\ref{fig:dynamics},\ref{fig:dynamics_various}}
        &\multicolumn{3}{c}{Fig.~\ref{fig:correlations}}\\
    \hline
      & $\xi_\Delta(\delta=0.3)$  & $\xi_\Delta(\delta=1.0)$ & $\xi_\Delta(\delta=3.0)$ &$\xi_\Delta$&$\xi_\Gamma$&$\xi_\Upsilon$&$\xi_\Theta$&$\xi_\Gamma$&$\xi_\Upsilon$&$\xi_\Theta$\\
     \hline
     KAM &3.0 & 4.0 &5.0 &4.0 &\multicolumn{3}{c|}{8.0} &\multicolumn{3}{c}{4.0}\\
     ABM &3.0  &3.0 &2.0 &3.0 &\multicolumn{3}{c|}{5.0} &\multicolumn{3}{c}{3.0}\\
     \hline
  \end{tabular}
\end{table*}
\endgroup

{\subsection{Determination of coarse-graining length}\label{sec:xi}
In this subsection, we explain how we determined the coarse-graining lengths $\xi_X(X\in\lbrace\Gamma,\Upsilon,\Theta,\Delta\rbrace)$ that {are} used for the analyses in the following sections (or already used in Figs.~\ref{fig:indicators_KAM}a and \ref{fig:indicators_ABM}a).
Since the coarse-graining lengths for the structural indicators are determined depending on the coarse-grained dynamic propensity field, we explain that for the dynamic propensity $\xi_\Delta$, first.

Coarse-grained structural indicators exhibit spatially smooth profiles as already presented in Figs.~\ref{fig:indicators_KAM}a and \ref{fig:indicators_ABM}a.
On the other hand, as shown in Figs.~S4 and S5 in the Supplemental Material, the bare dynamic propensity field without coarse-graining shows noisy profiles even within each mobile/immobile domain.
When we try to quantify the dynamic heterogeneity, we are interested in the meso-scale domain exhibited by the propensity field.
However, in the presence of these intra-domain noises, the estimation of correlation with structural indicators suffers from high-frequency modulations.
To exclude this unintentional underestimation of the correlation, we coarse-grained the propensity field as well.
To systematically determine the coarse-graining length {$\xi_X$}, we first measure the spatial correlation function of the dynamic propensity {$X =\Delta$}:
\begin{align}
C_\Delta(r)=\langle\delta\Delta(\boldsymbol{r}_i)\delta\Delta(\boldsymbol{r}_i+\boldsymbol{r})\rangle_{|\boldsymbol{r}|=r},\label{eq:sp_cr} 
\end{align} 
where $\delta\Delta\equiv\Delta-\bar{\Delta}$ is the deviation from the global average $\bar{\Delta}$, $r$ is the distance from the reference particle $i$, and $\langle\cdot\rangle_{|\boldsymbol{r}|=r}$ stands for the spherical average over particle pairs separated by a distance $r$.
Since we aim to smooth out the intra-domain noises here, we employ the decay length $r^\ast$ defined by $C_\Delta(r^\ast)\approx 1/e$ as the coarse-graining length $\xi_\Delta$.
In Fig.~\ref{fig:dynamics_spatial_corre}, we plot the measurement results of $C_\Delta$ at three different time scales for both KAM (panel a) and ABM (panel b).
For later convenience, in this study, we express the dynamic propensities at different time scales as functions of the mean intensity of the displacement (here, the displacement is cage-relative one. and the average is taken over the isoconfigurational samples and particles),$\delta(t)=\frac{1}{N}\sum_i^N{\Delta_i(t)}$, as $\Delta(\delta)$.
In Fig.~\ref{fig:dynamics_spatial_corre}, $C_\Delta$ at $\delta=0.3,1.0,3.0$ are shown.
These values of $\delta$ are expected to correspond to approximately $t\approx\tau_\alpha, 3\tau_\alpha,10\tau_\alpha$, where $\tau_\alpha$ is the $\alpha$ relaxation time~\cite{msd_alpha}.
We summarize the values of extracted coarse-graining length in Table~\ref{table:xi_values}.
Below, we use these values of $\xi_\Delta$ for $\Delta$ at these three time scales.

{We would like to stress that the coarse-graining of the dynamic propensity $\Delta$ introduced here seems not just an artificial operation but a physically reasonable one.
To show this, we prepared $100$ independent isoconfigurational samples and considered four different ensembles.
For three of the ensembles, we employed $N_s=30$, and completely different sets of samples are composed for each.
We call these ensembles $e_{30}^i (i\in\lbrace A,B,C\rbrace)$
Only in the forth ensemble, all $100$ samples are used ($N_s=100$): we call this $e_{100}$.
Because the number of samples is different, we expect that the ensemble $e_{100}$ should provide the statistically most reliable result.
In Fig.~\ref{fig:comparison_ensemble}, we compare the bare dynamic propensity fields and the coarse-grained ones obtained from these four ensembles (only results for the KA system are shown).
Although we see strong fluctuations among the bare fields of ensembles $e_{30}^i$ (Figs.~\ref{fig:comparison_ensemble}a-c), the coarse-grained fields of these ensembles (Figs.~\ref{fig:comparison_ensemble}e-g) appear highly similar.
Notably, the bare field of the ensemble $e_{100}$ (Fig.~\ref{fig:comparison_ensemble}d) is much smoother than the ones of ensembles $e_{30}^i$, and rather very similar to the coarse-grained fields (Figs.~\ref{fig:comparison_ensemble}e-g) of them.
These similarities can be quantitatively evaluated by measuring the Pearson's coefficient between two propensity fields of different ensembles.
We summarize the results in Table~\ref{table:comparison_ensemble}.
In this table, $C_{\Delta,\Delta}(e_\alpha,e_\beta)$, the correlations between the propensity fields of ensemble $e_i$ and $e_j$, are presented.
When we consider the coarse-grained field instead of the bare ones, we denote them as $\bar{e}_i$.
This table shows that the coarse-grained fields obtained from ensembles of $N_s=30$ are very close to each other ($C_{\Delta,\Delta}(\bar{e}_{30}^i,\bar{e}_{30}^j)=0.958$.
Here, regarding the correlation coefficient involving $e_{30}$, the average value over all combinations of $i$ and $j$ is shown, where $i,j\in\lbrace A,B,C\rbrace$ and $i\ne j$.) while the correlations between the bare fields are much smaller ($C_{\Delta,\Delta}(e_{30}^i,e_{30}^j)=0.769$).
This indicates that the bare $\Delta$ fields contain sample-dependent large intra-domain noises and our coarse-graining procedure indeed smooth away those unintentional sample-dependent noises as we desired.
Comparison between the results from ensembles with different values of $N_s$ further provides an important insight into the meaning of the coarse-graining of $\Delta$.
As expected, the correlation between non-coarse-grained and coarse-grained fields of $e_{100}$, namely $C_{\Delta,\Delta}(e_{100},\bar{e}_{100})=0.878$, is much larger than that of $e_{30}$, $C_{\Delta,\Delta}(e_{30},\bar{e}_{30})=0.820$, indicating that the coarse-grained field of the ensemble with $N_s=\infty$ would be identical to its bare field.
Moreover, we also mention that the correlation between coarse-grained ensembles $\bar{e}_{30}$ and $\bar{e}_{100}$ exhibits a very high value, $C_{\Delta,\Delta}(\bar{e}_{30},\bar{e}_{100})=0.985$.
All these results suggest that the coarse-graining of the dynamic propensity $\Delta$ is an important operation that allows to accurately estimate the ``genuine'' dynamic propensity field (that should be achieved in the limit of $N_s\to\infty$) from the numerical results with a finite $N_s$.}


\begin{figure*}[tb]
    \centering
    \includegraphics[width=\linewidth]{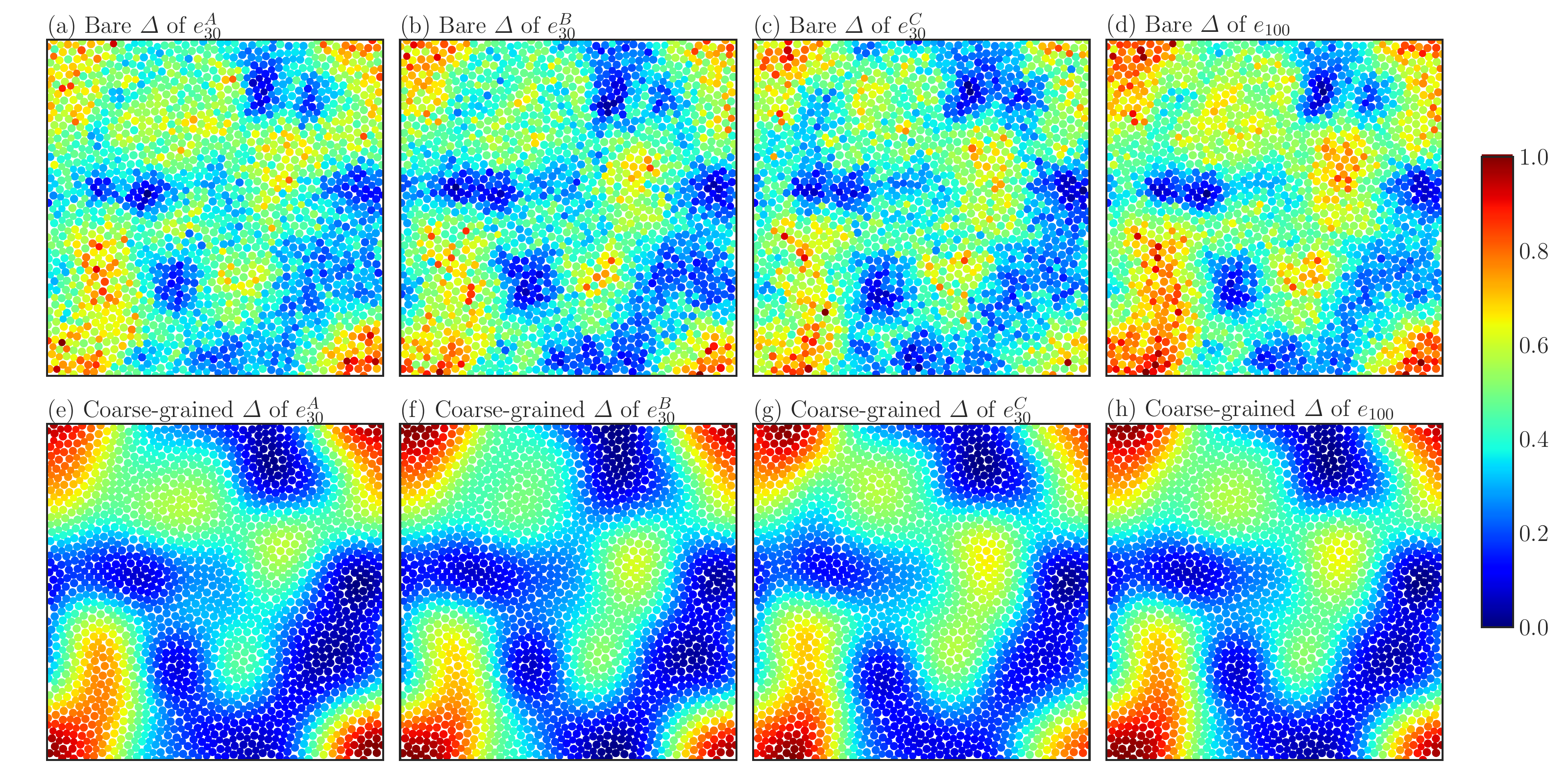}
    \caption{
Comparisons between the bare and the coarse-grained dynamic propensity fields $\Delta$ of different ensembles.
The results of $\Delta(1.0)$ of the KA system are shown. 
(a-c) The bare $\Delta$ fields of $e_{30}^i(i\in\lbrace A,B,C\rbrace)$, (d) the bare $\Delta$ field of $e_{100}$, (e-g) the coarse-grained $\Delta$ of $e_{30}^i$, and (h) the coarse-grained field of $e_{100}$. 
The coarse-graining length of $\xi_\Delta=4.0$ is employed as in Fig.~\ref{fig:indicators_KAM}e.
}
    \label{fig:comparison_ensemble}
\end{figure*}

\tabcolsep = 8pt
\begingroup
\renewcommand{\arraystretch}{1.5}
\begin{table*}[tb]
  \caption{Pearson's coefficients between $\Delta$ of different ensembles}
  \centering\label{table:comparison_ensemble}
  \begin{tabular}{ccccc}
  \hline
$C_{\Delta,\Delta}(e_{30}^i,e_{30}^j)$ &$C_{\Delta,\Delta}(\bar{e}_{30}^i,\bar{e}_{30}^j)$ &$C_{\Delta,\Delta}(e_{30},\bar{e}_{30})$  &$C_{\Delta,\Delta}(e_{100},\bar{e}_{100})$ &$C_{\Delta,\Delta}(\bar{e}_{30}, \bar{e}_{100})$\\
  \hline
0.769  &0.958 &0.820 &0.878 &0.985\\
 \hline
  \end{tabular}
\end{table*}
\endgroup

The coarse-graining lengths $\xi_\alpha$ for structural indicators $\alpha (\alpha\in\lbrace\Gamma,\Upsilon,\Theta\rbrace)$ are then determined in the same manner as the one in ref.~\cite{TTNatCommun}: the values that maximize the Pearson's correlation coefficient~\cite{Bapst2020,Paret2020,Boattini2020,Coslovich2022} between structural indicators and the dynamic propensity are chosen.
Here, as the dynamic propensity field, we employed the coarse-grained ones with $\xi_\Delta$ determined in the previous paragraph.
The determined values of $\xi_\alpha$ are summarized in Table~\ref{table:xi_values}.
See Supplemental Material for the detailed $\xi_\alpha$ dependence of the correlations.
In Figs.~\ref{fig:indicators_KAM}(b,c) and \ref{fig:indicators_ABM}(b,c), we show the visualization results of the coarse-grained Voronoi volume $\Upsilon$ and TT-OP $\Theta$ fields (the results for the same configurations as Figs.~\ref{fig:indicators_KAM}(a) and \ref{fig:indicators_ABM}(a)).
We stress that all panels present much larger domains than the size simply expected from the value of $\xi_\alpha$ (e.g., linear spanning of $2\xi_\alpha$).

As a reference, we also present the results without the coarse-graining of the dynamic propensity $\Delta$ (that is, results with $\xi_\Delta=0$) in Supplemental Material (Figs.~S4 and S5).
We note that, as shown in Figs.~S4 and S5, the consequences of the coarse-graining of $\Delta$ are fairly consistent with the bare $\Delta$ field.
}

\begin{figure}[tb]
    \centering
    \includegraphics[width=\linewidth]{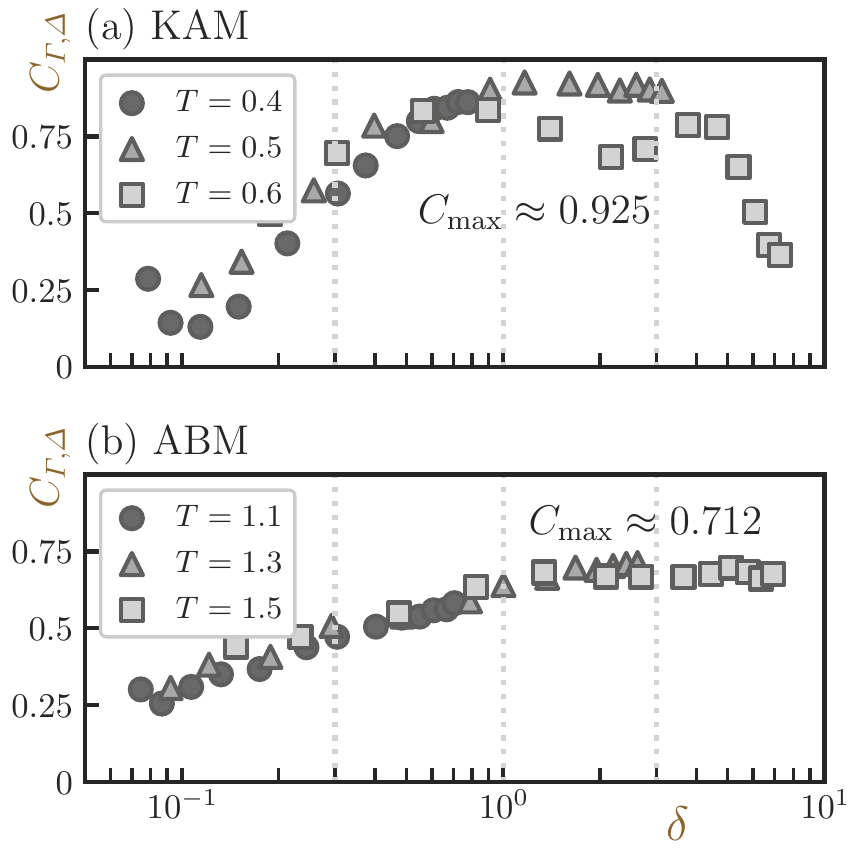}
    \caption{
    The ``time'' evolution of the correlation between Grad-CAM score $1-\Gamma$ and dynamic propensity $\Delta$ as a function of the intensity of the cage-relative displacement $\delta$.
    Each simulation was performed for $3\times 10^8$ steps, and $\delta(t)$ during those simulations is plotted on the abscissa for each plot.
    Different markers are used to distinguish different temperatures as shown in the legend.
    The vertical dotted lines indicate the delta values visualized in Figs~\ref{fig:indicators_KAM} and \ref{fig:indicators_ABM}.}
    \label{fig:dynamics}
\end{figure}
\subsection{Predictability of the dynamics}\label{sec:predict_dynamics}
Now we ask the following question: \emph{Are the extracted structures correlated with some material properties, for example, the dynamics?} To address this, we compared the Grad-CAM scores ($\Gamma$) with the dynamic propensity ($\Delta$). {Owing to the computational cost, we calculated the propensities only for the configurations shown in Figs.~\ref{fig:indicators_KAM} and \ref{fig:indicators_ABM} (only one configuration for each system). 
}

{For each configuration, we performed MD simulations with 30 different initial velocity distributions and calculated the dynamic propensity field $\Delta$ following the procedure summarized in Sec.~\ref{sec:propensity}.} 
Regarding the temperature during the measurement of the dynamics, we considered temperatures slightly above the glass transition point, $T^\ast$, (whose empirical definition is provided in the Supplemental Material; the obtained values are $T^\ast\approx 0.37$ for KAM and $T^\ast\approx 1.0$ for ABM) because we cannot expect any cage-breaking relaxational dynamics below $T^\ast$ within the computationally accessible time window. To investigate the possible temperature dependence of the dynamics, we performed simulations at temperatures up to approximately $1.5T^\ast$.
We stress again that although the initial velocities follow the specified temperatures (which are higher than the glass transition point $T^\ast$), the initial configurations are drawn from the sample at $T_{\rm G}=0.05$ (those shown in Figs.~\ref{fig:indicators_KAM} and \ref{fig:indicators_ABM}).
In Figs.~\ref{fig:indicators_KAM}(d-f) and \ref{fig:indicators_ABM}(d-f), the propensity $\Delta$ at $\delta\approx 0.3, 1.0, 3.0$ are shown.
Note again that we express the time-dependence indirectly via $\delta$, the mean intensity of the cage-relative displacement, and these values of $\delta$ correspond roughly to $t\approx\tau_\alpha, 3\tau_\alpha, 10\tau_\alpha$ respectively.
Interestingly, there is agreement between $1-\Gamma$ and $\Delta$ at long times ($\Delta(1.0)$ and $\Delta(3.0)$) for both systems.

To quantify the correlation between $\Gamma$ and $\Delta$ further, we calculated the Pearson's correlation coefficients, $C_{\Gamma,\Delta}$. 
{Although the coarse-graining length of $\Delta$ is dependent on the value of $\delta$ (i.e., the time scale), for simplicity, we employed a fixed value for each system here (see Table~\ref{table:xi_values}).}
The results are presented in Fig.~\ref{fig:dynamics}.
In this plot, the time dependence is indirectly reflected by the value of $\delta$. 
Such a presentation allows us to compare the correlation between the dynamics (at different temperatures) and the static structure directly, thus ruling out the effect of the nontrivial dependence on time. 
Note that the correlation between $1-\Gamma$ and $\Delta$ is quantified, not $\Gamma$, in agreement with the visualization. 
From Fig.~\ref{fig:dynamics}, we observe several striking consequences. 
First, $C_{\Gamma, \Delta}$ rises in accordance with the increase in $\delta$, reaching its maximum value at $\delta^>\approx 1$ in both the KAM and ABM. 
This indicates that the structures extracted by our method are responsible for the dynamics at a longer time scale than the $\alpha$ relaxation (note that these are nonequilibrium aging dynamics and not intra-metabasin equilibrium relaxation). 
Secondly, the change in the correlation $C_{\Gamma, \Delta}$ is nonmonotonic in the KAM system and starts decreasing for $\delta\ge \delta^\ast$, while plateauing for $\delta\ge \delta^\ast$ in the ABM system.
These results indicate that the specified characteristic ``well-ordered'' clusters are transient in the KAM system, whereas they seem very stable within the time window of our calculation in the ABM system.
Thirdly, the maximum correlation, $C_{\rm max}$, reaches very high values in both systems: {0.925 and 0.712 in the KAM and ABM}, respectively.
The predictability of the dynamics is surprising because our method does not require any information about the dynamics during the training process; thus, the computational cost for both the training and the sample-preparation part is low. 
Finally, the results of different $T$ follow a single master curve.  
This result confirms the fact that the dynamics are indeed governed by the static ``glass structures,'' at least in the temperature regime under study and concerns nonequilibrium aging dynamics.

\subsection{Interpretable structural indicators}
In this section, we measure two distinct local multibody structural indicators to interpret the Grad-CAM score $\Gamma$. Because these indicators are handmade, we can take advantage of their interpretable nature. In the Supplemental Material, we present the two-body correlation function $g(r)$ for reference (Fig.~S7).
We again stress that all the particle-based indicators, including $\Delta$, were coarse-grained and further normalized to the interval $[0,1]$. 

{\it Voronoi volume.~--~}
The Voronoi volume $\Upsilon$ is the first interpretable local multiparticle structural indicator. In this subsection, we briefly explain the obtained $\Upsilon$ values for the KAM and ABM systems. 
A typical result for the KAM system is shown in Fig.~\ref{fig:indicators_KAM}(b) in which particles with small values of $\Upsilon_i$ ($<0.4$) appear dominant.
Fig.~\ref{fig:indicators_ABM}(b) presents the results for the ABM system, in which a large portion of particles exhibit relatively high values of $\Upsilon_i$ ($>0.4$). 
Note that $1-\Upsilon$ is visualized in Fig.~\ref{fig:indicators_ABM}(b).

These distinct trends are derived entirely from the difference in the set of interaction parameters ($\epsilon_{ij}$ and $\sigma_{ij}$) and the number ratio of the particle species ($N_A/N_B$). 
For instance, in the KAM system, the interaction energy is most stable when different species are in contact, and the interaction range is also the shortest in this situation (see Table~\ref{table:parameters}). Therefore, small values of $\Upsilon_i$ are energetically favored in the KAM.
In the ABM system, on the other hand, because the area occupied by particles {A} is almost half that of particles {B (the area fraction is $1:1.96$)}, the region with a large Voronoi volume (corresponding to particles B) tends to be slightly dominant. 
Because samples with a very low temperature $T=0.05$ are shown in Figs.~\ref{fig:indicators_KAM} and \ref{fig:indicators_ABM}, the structurally low-energy states are expected to be more probable. 
We also note that the small value of $\Upsilon_i$ does not necessarily mean that the local structure around particle $i$ is highly ordered, as is evident in the case of the KAM.

{\it Tong--Tanaka order parameter.~--~}
The second interpretable structural indicator is the TT-OP $\Theta$. 
As mentioned in Sec.~\ref{sec:ttop}, of the various locally defined structural indicators reported to date, {TT-OP captures the dynamical behavior of many classes of glasses very well, especially universally.} 
The characteristic structures of glasses in terms of the TT-OP are specified by small values of $\Theta_i$, which means that the local structure is highly ordered. 

The results for the KAM system are shown in Fig.~\ref{fig:indicators_KAM}(c). 
{We note that because of the nonadditive nature of the potential parameters, defining a reference three-particle ideal configurational angle $\theta^1$ in the case of the KAM is nontrivial.} In this study, we employed the definition using the additive assumption ($\sigma_{AA}=1.0, \sigma_{BB}=0.88$, and $\sigma_{AB}={(\sigma_{AA}+\sigma_{BB})/2=}0.94$) rather than the parameters used in the simulations because the reference structure is easier to interpret with this {additive} assumption. 
Surprisingly, the coarse-grained $\Theta$ field looks very similar to other indicators and appears correlated to the long-time propensity field.
The results for the ABM system are shown in Fig.~\ref{fig:indicators_ABM}(c). In contrast to the findings of Tong and Tanaka~\cite{TTPRX,TTNatCommun}, the structure is not well developed, and the spatial modulation is much smaller than that in the KAM: intermediate values are system-spanning.
This is likely because our samples were generated by quenching at a fixed cooling rate and, thus, were not well annealed. We expect these samples to exhibit aging behavior.

\subsection{Correlations between different indicators}\label{sec:comparison}
\begin{figure}[tb]
    \centering
    \includegraphics[width=\linewidth]{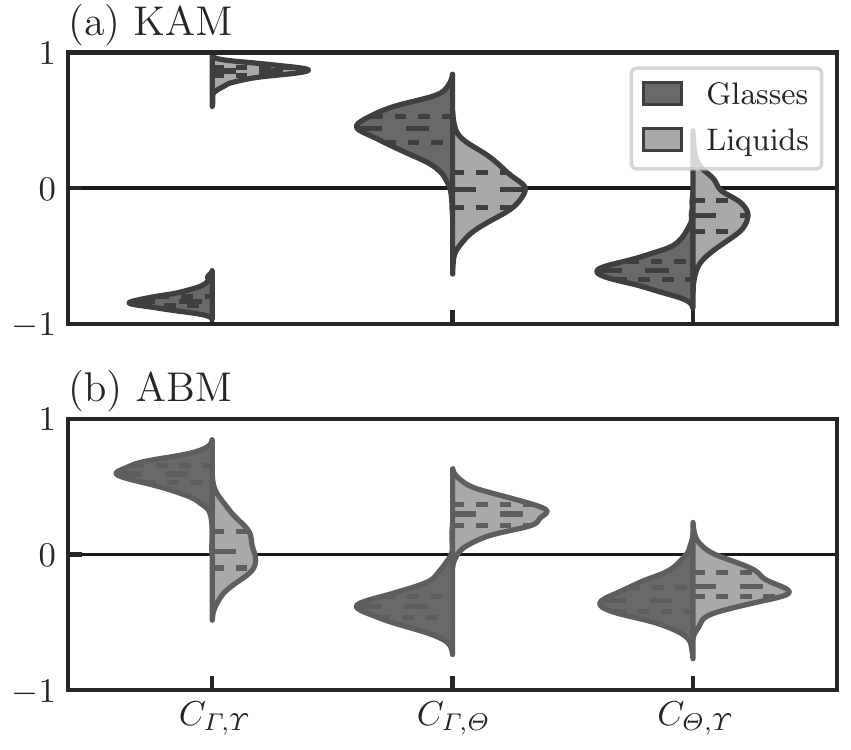}
    \caption{
    Violin plots of the Pearson's correlation coefficients between distinct structural indicators. 
    Results for the (a) ABM and (b) KAM systems.
    The dark and light gray parts represent the results for glasses liquids, respectively, as shown in the legend.
    The dashed lines indicate the quartiles.
    }
    \label{fig:correlations}
\end{figure}

To interpret the Grad-CAM scores ($\Gamma$) obtained by our method, we further calculated the Pearson's correlation coefficients between different indicators for both the KAM and ABM. The results are summarized in Fig.~\ref{fig:correlations}. In this figure, we show violin plots of the coefficients between different indicators ($\Gamma, \Upsilon$ and $\Theta$) calculated using {600} samples for each case (KAM or ABM and glasses or liquids). 
{In this subsection, for clarity, we employ different values of $\xi_\alpha$ from those used in other subsections as shown in Table~\ref{table:xi_values}.
The change in the value of $\xi_\alpha$ does not change the qualitative discussion here, while the distributions (those are plotted in Fig.~\ref{fig:correlations}) become broader, and thus the differences between them are less pronounced when $\xi_\alpha$ increases.}

We call the correlation coefficients between the Grad-CAM score and the Voronoi volume $C_{\Gamma,\Upsilon}$ and define those between different pairs in a similar manner: $C_{\Gamma,\Theta}$ and $C_{\Theta,\Upsilon}$. Although the results below are mostly those for the glass configurations only, we also mention the results for the liquids when we discuss their differences from those of the glass samples. Finally, we stress that, in the main text, all correlations are based on the Pearson's definition. As presented in the Supplemental Material, however, we also obtained semi-quantitatively consistent results using Spearman's definition. 
Below, we explain the results for the ABM and KAM systems.

{\it ABM.~--~}
In the ABM system, $C_{\Gamma,\Upsilon}$ (the Grad-CAM score vs. the Voronoi volume) is the largest in terms of the intensity (Fig.~\ref{fig:correlations}b), and its average is an intermediate positive value: $\bar{C}_{\Gamma,\Upsilon}^{\rm G}=0.585$, where the bar represents the average over samples and the superscript G indicates that only the glass samples are considered (see Table~S2 in the Supplemental Material for the summary of the average and the standard deviation of the correlation coefficients). This means that structures with large local volumes are judged to be characteristic of the glass.
This behavior is consistent with the TT-OP: both $C_{\Gamma,\Theta}$ and $C_{\Theta,\Upsilon}$ (those for the Grad-CAM score vs. the TT-OP and the TT-OP vs. the Voronoi volume, respectively) are negative, with the intensities being slightly smaller ($\bar{C}_{\Gamma,\Theta}^{\rm G}=-0.378$ and $\bar{C}_{\Theta,\Upsilon}^{\rm G}=-0.327$), meaning that structures with large values of the Voronoi volume tend to be more ordered.

{However, importantly, the difference between glasses and liquids is most evidently quantified by $C_{\Gamma,\Theta}$, which becomes almost completely negative for glasses but positive for liquids.
On the other hand, the probability distribution of $C_{\Gamma,\Upsilon}$ shows a large overlap between glasses and liquids; moreover, in the case of liquids, the distribution is centered around zero, indicating that the Voronoi-volume-like aspect of the Grad-CAM score is likely unable to distinguish glasses and liquids accurately. Therefore, our method seems to rely on structures that are qualitatively consistent with the TT-OP rather than on the Voronoi volume when a decision is made, although $\Upsilon$ is closer to $\Gamma$ than $\Theta$ in terms of the correlation for glass samples, as mentioned above ($\bar{C}^{\rm G}_{\Gamma,\Upsilon}>\bar{C}^{\rm G}_{\Gamma,\Theta}$). 
We stress that {it has been shown} that the TT-OP can extract the characteristic structures associated with the dynamics in binary additive glass formers~\cite{TTNatCommun}, and our results seem consistent with this.}

{\it KAM.~--~}
In the case of the KAM system, $C_{\Gamma,\Upsilon}$ and $C_{\Gamma,\Theta}$ for glasses are negative and positive, respectively, at odds with the results for the ABM (Fig.~\ref{fig:correlations}a).
Such a qualitative difference indicates that the structures extracted by our method respect the details of the systems. It should also be noted that, from the perspective of the intensity of the correlation coefficients, $C_{\Gamma,\Upsilon}$ is significantly larger than $C_{\Gamma,\Theta}$, and only $C_{\Gamma,\Upsilon}$ exhibits a clear difference in the signs between the results for glasses and liquids. This is another qualitative difference from the ABM system, where the difference in sign is evident for $C_{\Gamma,\Theta}$. 
However, although the TT-OP is a good descriptor for the ABM, as presented above, it is unlikely to characterize the properties of KAM systems. Thus, our method regards structures with high $C_{\Gamma,\Upsilon}$ as characteristic while $C_{\Gamma,\Theta}$ is small. In particular, the intensity of $C_{\Gamma,\Theta}$ is lower than that of $C_{\Theta,\Upsilon}$ ($\bar{C}_{\Gamma,\Theta}^{\rm G}=0.426$, $\bar{C}_{\Theta,\Upsilon}^{\rm G}=-0.597$), suggesting that our method attempts to avoid correlation with the TT-OP selectively.

{\it Summary of this subsection.~--~}
Interestingly, although we could interpret the Grad-CAM scores in terms of other conventional indicators (the Voronoi volume and the TT-OP) to some extent in both the KAM and ABM, the correlations are not perfect, and Grad-CAM seems to blend different indicators in an ``appropriate'' manner. In particular, we emphasize that the precise recipe of such blending is obviously dependent on the system details. 
Therefore, it would be meaningful to regress the obtained Grad-CAM score field $\Gamma$ symbolically to achieve a fuller interpretation using recently invented methods~\cite{Brunton2016,Rudy2017,Schoenholz2017}.

\begin{figure}[tbh]
    \centering
    \includegraphics[width=\linewidth]{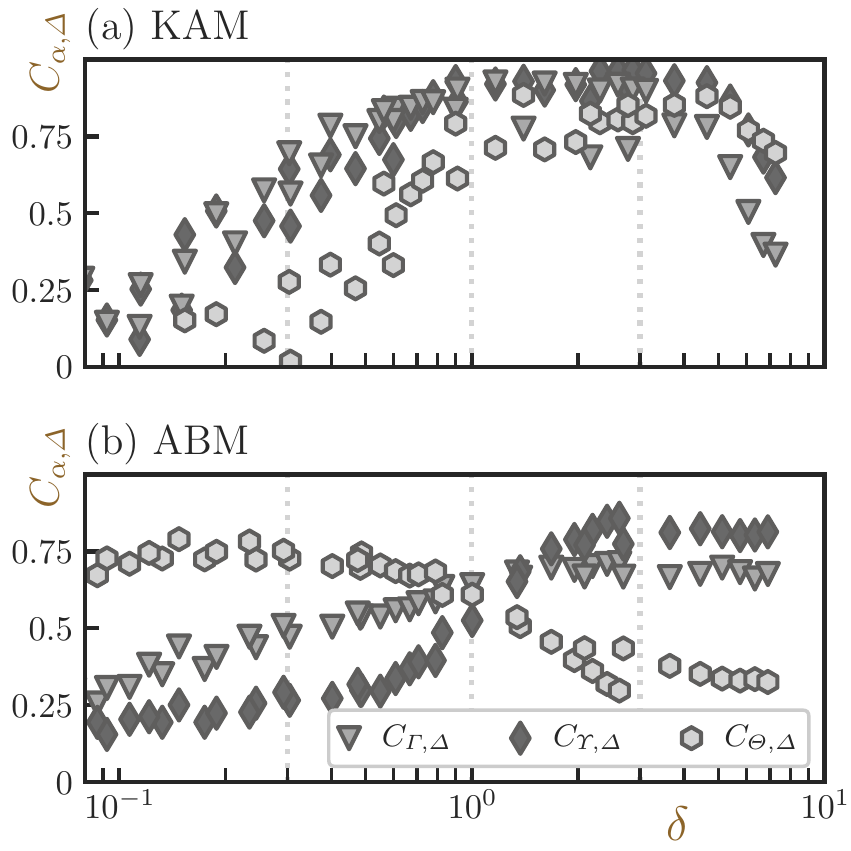}
    \caption{
    The evolution of the correlation between structural indicators (Grad-CAM score $\Gamma$, Voronoi volume $\Upsilon$, and TT-OP $\Theta$) and dynamic propensity $\Delta$ (or $1-\Delta$ depending on the target indicator and system) as a function of the mean intensity of the cage-relative displacement $\delta$.
    The meaning of the abscissa is the same as the one in Fig.~\ref{fig:dynamics}.
    Different markers distinguish different indicators, as shown in the legend.
    Data for all temperatures are plotted without distinction.
    The vertical dotted lines indicate the delta values visualized in Figs~\ref{fig:indicators_KAM} and \ref{fig:indicators_ABM}.}
    \label{fig:dynamics_various}
\end{figure}
\tabcolsep = 14pt
\begingroup
\renewcommand{\arraystretch}{1.5}
\begin{table}[tb]
  \caption{Maximum values of correlations and their locations}
  \centering\label{table:dynamics_various}
  \begin{tabular}{c|cc|cc}
  \hline
    &\multicolumn{2}{c|}{KAM}    &\multicolumn{2}{c}{ABM}\\
    \hline
     &$C^{\rm max}_{\alpha,\Delta}$ & $\delta^\ast$ &$C^{\rm max}_{\alpha,\Delta}$ & $\delta^\ast$\\
    \hline 
    $C_{\Gamma,\Delta}$ &$0.925$ &$1.167$ &$0.712$ &$2.620$\\
    $C_{\Upsilon,\Delta}$ &${\bf 0.966}$ &$2.860$ &${\bf 0.858}$ &$2.620$\\
    $C_{\Theta,\Delta}$ &$0.885$ &$1.402$ &$0.790$ &$0.148$\\
    \hline
    \multicolumn{5}{c}{(The best $C_{\alpha,\Delta}^{\rm max}$ is shown in bold letters for each system)}
  \end{tabular}
\end{table}
\endgroup

{\subsection{Dynamics vs. other indicators}\label{sec:dynamics_various}
In Sec.~\ref{sec:predict_dynamics}, we studied the predictability of the Grad-CAM score $\Gamma$ with respect to the dynamics $\Delta$ by measuring the correlation coefficient between them.
In this subsection, we further investigate the correlation between dynamics and other indicators, namely, the Voronoi volume $\Upsilon$ and the TT-OP $\Theta$. 
Fig.~\ref{fig:dynamics_various} presents the correlation coefficient between the dynamic propensity and the structural indicators.
Note that, in this subsection, when calculating the correlation $C_{\alpha, \Delta}$, we sometimes use $1-\alpha$ instead of $\alpha$ to 
obtain a positive value (the choices obey those in Figs.~\ref{fig:indicators_KAM} and \ref{fig:indicators_ABM}).
We explain the results for $\Upsilon$ and $\Theta$ one by one below.

{\it~Voronoi volume.~--~}
The time evolution of the correlation coefficient between the Voronoi volume $\Upsilon$ and the dynamic propensity $\Delta$, $C_{\Upsilon,\Delta}$, is quite similar to that of $C_{\Gamma,\Delta}$:
it changes non-monotonically (reaches the maximum at $\delta^\ast> 1.0$ and then starts decreasing) in the KAM system, and grows monotonically as a function of $\delta$ and saturates at $\delta^\ast>1.0$ in the ABM system.
The maximum correlation $C_{\alpha,\Delta}^{\rm max}$ and the value of $\delta^\ast$ at which $C_{\alpha,\Delta}(\delta^\ast)=C_{\alpha,\Delta}^{\rm max}$ holds are summarized in Table~\ref{table:dynamics_various}.
Here, the subscript $\alpha\in\lbrace\Gamma,\Upsilon,\Theta \rbrace$ distinguishes the indicator of interest.
The maximum $C_{\Upsilon,\Delta}^{\rm max}$ reaches high values in both systems: 0.966 in the KAM and 0.858 in the ABM (to obtain a positive value, we employed $1-\Upsilon$ for the ABM).
{These values are higher than $C_{\Gamma,\Delta}^{\rm max}$ in both systems.}
{This is a quite unanticipated consequence since the predictability of the free volumes with respect to the dynamic propensity has been negated previously~\cite{Widmer-Cooper2005,Widmer-Cooper2006}. 
On the other hand, it has been reported that the local potential field is strongly correlated with the dynamics when coarse-grained~\cite{Donati1999,Matharoo2006,Berthier2007}. 
Since both the local potential and the Voronoi volume detect the metric-based information of the local packing, we expect them to possess qualitatively similar information.
Therefore, it is possible that the good predictability of the Voronoi volume is a result of the coarse-graining.
To draw a decisive conclusion, however, a thorough investigation using the same setup as that in~\cite{Widmer-Cooper2005,Widmer-Cooper2006} is required.}

{\it~Tong-Tanaka order parameter.~--~}
Interestingly, the correlation between the TT-OP and the dynamic propensity, $C_{\Theta,\Delta}$, reaches high values: 0.885 and 0.790, respectively, in the KAM and ABM systems ($1-\Theta$ and $\Theta$ were employed).
It is unexpected that the TT-OP provides a good predictability of the dynamics even in the KAM (the correlation is higher in the KAM than in the ABM).
This good predictability is achieved maybe because we employed the additive convention of the reference angle $\theta^2_{ij}$ or because we focused on the non-equilibrium aging dynamics.
We also note that the high correlation with dynamics is observed only for long-time regimes, and the correlation is very low around the $\alpha$ relaxation regime ($\delta\approx 0.3$).
Further comprehensive investigations are necessary to identify the cause of the unexpectedly high predictability.

In the KAM system, the qualitative trend is the same as those for correlations of other indicators ($C_{\Gamma.\Delta}$ and $C_{\Upsilon,\Delta}$): it starts from a small value and follows an upward convex curve.
In the ABM system, in contrast, the time evolution of $C_{\Theta,\Delta}$ is qualitatively different from those of $C_{\Gamma,\Delta}$ or $C_{\Upsilon,\Delta}$. 
It is high even at the early-stage small $\delta$ regime and changes in a non-monotonic manner with the increase in $\delta$: it increases only a little bit, reaches the maximum value at $\delta^\ast\approx 0.148$, remains almost at the same level, and then starts decreasing.

{\it Summary of this subsection.~--~}
To summarize, first, the Voronoi volume has the largest correlation with the dynamics in both KAM and ABM systems, in terms of the $C_{\alpha,\Delta}^{\rm max}$.
Regarding the comparison between the Grad-CAM score and the TT-OP, the latter shows a stronger correlation in the ABM system (note again that the TT-OP is known to be a good descriptor of the dynamics in the ABM) while the former outperforms in the KAM.

The results presented in this article indicate that the characteristic structures extracted by the Grad-CAM capture information consistent with that of other coarse-grained structural indicators proposed in previous works~\cite{Boattini2020,Paret2020} in the sense that all structures are correlated with the dynamic propensity to some extent.
However, we do observe clear differences between the correlation coefficient for the TT-OP and the other two indicators (the Voronoi volume and the Grad-CAM score), particularly in the ABM: although $C_{\Gamma,\Delta}$ and $C_{\Upsilon,\Delta}$ reach their maximum values at $\delta> 1.0$, which corresponds to the longer time scales than $\alpha$ relaxation time, only $C_{\Theta,\Delta}$ exhibits clearly smaller values of $\delta^\ast$ (Table~\ref{table:dynamics_various}).
This may suggest that the structures specified by the TT-OP and those identified by the other indicators signal qualitatively different aspects of heterogeneous dynamics.
Indeed, while the TT-OP focuses on angular information, the other two indicators take into account the whole structural information.
}

\begin{figure}[tb]
    \centering
    \includegraphics[width=\linewidth]{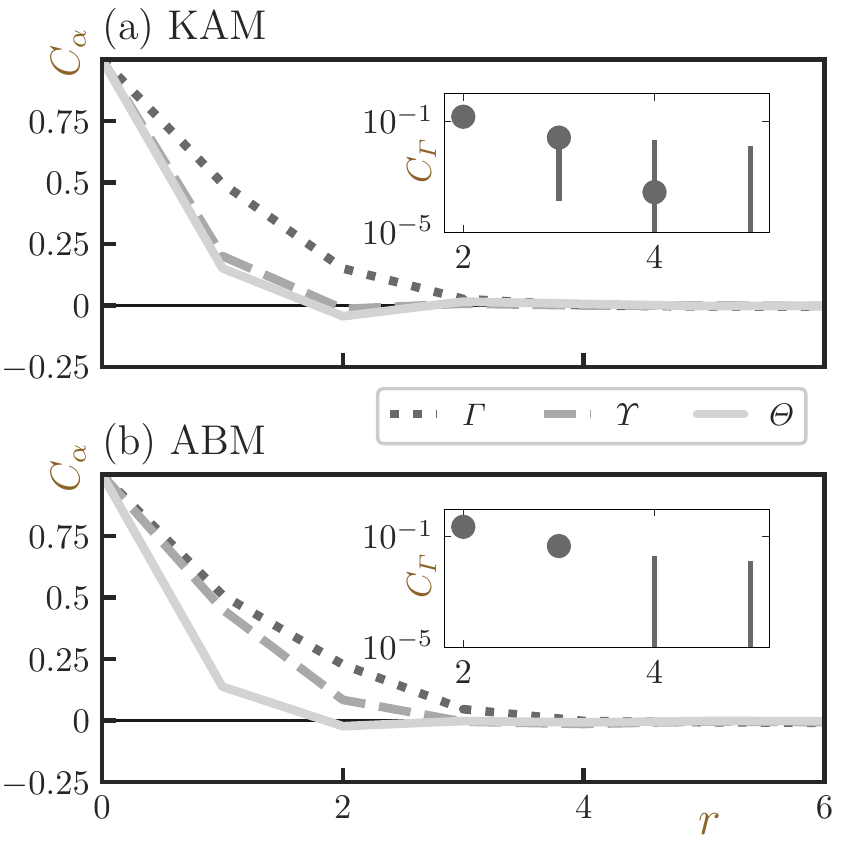}
    \caption{
{The spatial correlation functions of structural indicators $\alpha (\alpha\in\lbrace\Gamma, \Upsilon, \Theta\rbrace)$, $C_\alpha$, as a function of the distance from the reference particle $r$.
Results for (a) KAM, (b) ABM systems.
Different line styles distinguish different indicators as shown in the legend.
In the insets of both panels, the magnified images of $C_\Gamma$ in the vicinity of $C_\Gamma=0$ with error bars.
In these insets, the vertical axes are in the logarithmic scale.
Although the values of $C_\Gamma$ are negative when markers are missing, the error bars are crossing $C_\Gamma=0$.
}
}
    \label{fig:spatial_corre}
\end{figure}

{
\subsection{Do the coarse-graining lengths have a structural origin?}
Thus far, we have shown that, coarse-grained with proper choices of $\xi_\alpha$, structural indicators show strong correlations with the dynamics $\Delta$.
In this subsection, we discuss whether we can find the structural origin of these ``proper'' coarse-graining lengths.
To this end, we measured the same spatial correlation functions as the one in Eq.~\ref{eq:sp_cr} for structural indicators.
The results are presented in Fig.~\ref{fig:spatial_corre}.
We can first tell, from this figure, that $\Upsilon$ and $\Theta$ decay very fast to $C_\alpha=0$ (particularly in the case of KAM).
In contrast, $\Gamma$ decays relatively slowly in both KAM and ABM systems: $C_\Gamma$ reaches zero at $r_\Gamma\approx 3.0-4.0$.
This indicator-dependence of the correlation length is evident in the visualizations of bare fields shown in Figs.~S6 and S7 in the Supplemental Material.
Results in Fig.~\ref{fig:spatial_corre} suggest that relying on $\Gamma$, and we can extract relatively long purely structurally-based correlation length.
Note that, here, we define the correlation length as the one where $C_\Gamma$ becomes zero, not $1/e$.
We employed this choice because, unlike the case of $\Delta$ for which we introduced the coarse-graining to smooth out the intra-domain noises, we are interested in the average size of the domains as discussed below.

In both systems (KAM and ABM), the length scale $r_\Gamma$ defined here as $C_\Gamma(r_\Gamma)=0$ is roughly half of the optimal coarse-graining length $\xi_\alpha$ used in the aforementioned analyses ($\xi_\alpha=8.0$ for KAM and $\xi_\alpha=5.0$ for ABM).
Because $r_\Gamma$ is expected to correspond to the average domain size, the length scale $\xi_\alpha\approx 2r_\Gamma$ corresponds to the average distance between domains with the same sign of $\delta\Gamma\equiv \Gamma-\bar{\Gamma}$.
This correspondence suggests that the dynamical domains become larger than the static ones due to collective excitations of nearby mobile domains (that are expected to be indicated by low values of $\Gamma$). 
This picture is consistent with a recent theoretical work~\cite{Ozawa2022arxiv} where the dynamic heterogeneity was explained by collective behaviors of local instabilities (that correspond to mobile domains) via elastic interactions.

To further give a concrete interpretation of the coarse-graining length on a purely structural basis, we must understand the interactions between domains via the elastic field.
To provide useful data for such an exploration of new understandings, it is important to perform a comprehensive measurement of $\Gamma$ for equilibrated well-annealed low-temperature samples.
We leave this problem as a future work.
}


{\subsection{Relation to recent works}
In the closing remarks for this section, we discuss the relation of our work to several recent works using machine learning-based methods.
Recently, much effort has been dedicated to challenges in explaining the heterogeneous slow dynamics of glasses from a purely structural perspective.
For instance, in refs.~\cite{Bapst2020,Hayato}, supervised learning of graph neural networks was performed with information on dynamics as part of the training data.
The trained networks succeeded in predicting the long-time glassy dynamics of the KAM system at very low temperatures (the lowest value $T=0.44$ is comparable to the mode coupling transition point $T_{\rm MCT}=0.435$) solely from static structures with high precision (the correlation coefficient exceeds 0.6).
As examples of unsupervised approaches, refs.~\cite{Boattini2020,Paret2020} similarly tried to extract characteristic structures of glasses from static configurations.
In these studies~\cite{Boattini2020,Paret2020}, information from dynamics was not used, even in the training stage, similarly to our method.
The major difference to our method was that only glass configurations were provided during the training stage~\cite{Boattini2020,Paret2020}; the liquid samples were absent.
Surprisingly, the obtained structures were well-correlated with the long-time dynamics, particularly the dynamic propensity at approximately the $\alpha$ relaxation time.
{For the KAM system, the correlation coefficient reaches around 0.4 and 0.7 in refs.~\cite{Boattini2020,Paret2020} respectively.}

{Because} our method similarly does not require any information about the dynamics, we can say that it is also unsupervised in regard to dynamics prediction. {Accordingly, it is non-trivial and interesting that} the structures extracted using our method exhibit a strong correlation with the long-time heterogeneous dynamics at a longer time than the $\alpha$ relaxation time, as was the case for methods in refs.~\cite{Boattini2020,Paret2020}.
This implies that the characteristic structures governing the relaxational dynamics are extracted in a similar way, whether we try to identify the structural difference between glasses and liquids (our approach) or specify structurally distinct parts from the glass sample only (approaches in refs.~\cite{Boattini2020,Paret2020}).
This may support the view that the characteristic glass structures, if exist, grow gradually from completely random liquid configurations as the temperature decreases.
It would be very meaningful to investigate the similarity of structures extracted by different machine learning methods.

We also note the quantitative difference in the predictability of different machine-learning methods with respect to the dynamics.
Although we cannot make direct quantitative comparisons because of the varied setups in the references, our method provides the highest-level performance in terms of the simple correlation coefficient between extracted characteristic structures and long-time dynamic propensity: for our 2D ABM and KAM system, the correlation between the Grad-CAM score and dynamics reached approximately 70\% and 90\%, respectively.
}
\section{Summary and overview}\label{sec:summary}
In this work, we proposed a method to extract the characteristic structures of amorphous systems solely from a couple of classes of static random configurations by means of classification with a CNN and quantification of the grounds for classification using Grad-CAM.
We applied the proposed method to two qualitatively different binary glass-forming mixtures, viz. the ABM and KAM, and showed that our method could automatically extract the system-detail-dependent mesoscopic characteristic structures of glasses. The proposed method has three outstanding features. First, our method can extract characteristic structures solely from the instantaneous static structures without any information about the dynamics.  
Second, the extracted characteristic structures are system-dependent; in other words, our method automatically identifies the tailor-made structural indicator suitable for each distinct system. 
Finally, the extracted structures are strongly correlated with the dynamic propensity.
The time evolution of the correlation reveals that the characteristic structure is closely related to the dynamics at a longer time scale than that for the $\alpha$ relaxation, where the mean intensity of the cage-relative displacement is of the order of unity: $\delta\gtrapprox 1$. Moreover, such a correlation is robust over a wide range of temperatures, at least in the range $T^\ast\le T\le 1.5T^\ast$. 
{We again stress that, unlike in the previously reported studies, our dynamic propensity quantifies the nonequilibrium {aging processes}, not the intra-metabasin equilibrium relaxational dynamics, and is coarse-grained.}

In addition, we discuss several future research directions. First, we should conduct a similar analysis using well-annealed glass configurations because the equilibrium dynamics are important to understand the properties of glasses more deeply. In particular, it is challenging to determine whether the characteristic structures that our method extracts for equilibrium glass configurations are correlated with the equilibrium dynamics. Moreover, our method can find the characteristic structures from the static configurations alone, even when the microscopic physical quantity that characterizes the two classes (e.g., the dynamic propensity) is not available, as long as the two different classes are defined, for example, by specifying macroscopic quantities such as the temperature.
Therefore, it allows us to directly ask, for example, whether we observe any structural differences between normal and ultrastable glasses~\cite{Berthier2017,Khomenko2020}, for which the dynamical properties are numerically intractable.
Because ref.~\cite{Wang2019} reported that the stability of glass samples is structurally reflected by the density of the quasi-localized vibrational (QLV) modes, it would be interesting to see if Grad-CAM also quantifies the QLV modes or highlights completely different structures. Similarly, it has been shown that the structural difference between instantaneous configurations under different external shear speeds is quantified by the density of the imaginary normal modes~\cite{Oyama2021}.
Our method is also applicable in these situations.

In general, we expect that the extracted structures are dependent on the precise setup of the classification problem, such as the temperature choice for each class. 
It would be interesting to compare the characteristic glass structures obtained from different reference high temperatures. 
Further, such structures could be sensitive to the details of the protocols, for example, the network architecture or the number of epochs. 
The investigation of the effects of these factors would be valuable.

{
The authors thank Satoshi Koide, Takenobu Nakamura, Hayato Shiba, and Ludovic Berthier for fruitful discussions and comments. This work was financially supported by the JST FOREST Program (Grant No. JPMJFR212T) and JSPS KAKENHI (Grant Nos. 22H04472, 20H05157, 20H00128, 19K03767, 18H01188).}
\bibliography{main}

\end{document}


\title{
What Do Deep Neural Networks Find in Random Structures? --- supplemental material
}

\author{Norihiro Oyama}
\email{Norihiro.Oyama.vb@mosk.tytlabs.co.jp}
\affiliation{Toyota Central R\&D Labs, Inc., Bunkyo-ku, Tokyo 112-0004, Japan}

\author{Shihori Koyama}
\affiliation{Toyota Central R\&D Labs, Inc., Bunkyo-ku, Tokyo 112-0004, Japan}

\author{Takeshi Kawasaki}
\affiliation{Department of Physics, Nagoya University, Nagoya 464-8602, Japan}

\maketitle

\section{Technical details}

\subsection{Pre-processing of input data}
Because the particle configuration data are not compatible with the CNN as they are, we need to process them before feeding to the network.
In this article, we have gridized the particle configuration data $\rho(\boldsymbol{r})=\sum_i^N\delta(\boldsymbol{r}-\boldsymbol{r}_i)$ using a mapping operator ${\cal M}$ whose operation on the grid point specified by $(k,l)$ is defined using a simple Gaussian kernel $f(x)=\exp(-x^2/\sigma^2)$ ($\sigma$ is the width of the kernel) as:
\begin{align}
\tilde{\rho}(\tilde{\boldsymbol{r}}_{k,l})&={\cal M}_{k,l}(\rho(\boldsymbol{r})),\\
&=\sum_{i\in\partial(k,l)}f(|\tilde{\boldsymbol{r}}_{k,l}-\boldsymbol{r}_i|),\label{eq:filter}
\end{align}
{where $\tilde{\boldsymbol{r}}_{k,l}$ is the position vector of the grid point specified by the indices $k$ (for $x$-direction) and $j$( $y$-direction) and $\tilde{\rho}({\boldsymbol{r}}_{k,l})$ is the grid-based density field (in concert with the main text, the tilde symbol is used to denote the grid-based variables).
The kernel width $\sigma$ depends on the "size" of the particle $i$ ($\sigma_{ii}$) as  $\sigma=a\sigma_{ii}$, where the factor $a=L/\sqrt{0.64N}$ is of the order unity and $L$ is the linear dimension of the system.}
The term $\partial(k,l)$ appearing in Eq.~\ref{eq:filter} gives the list of particles that lie within the cutoff distance $r_{\rm c}^{\cal M}=2.5\sigma_{ii}$ (fixed for both ABM and KAM) from the grid point $(k,l)$. 
Note that when the calculated density field $\tilde\rho$ is fed to the CNN as the input, they are normalized to the interval of $[0,1]$ for the computational efficiency, by simply subtracting the minimum value followed by the division by the maximum value.

Inversely, after the calculation of the grid-based Grad-CAM score $\tilde{L}^C$, we would like to put them back to particles.
This is similarly achieved by the inverse operator ${\cal M}^{-1}$ whose operation on the particle $i$ is defined as:
\begin{align}
    \Gamma_i&={\cal M}_i^{-1}(\tilde{L}^C),\\
    &=\sum_{(k,l)\in\partial i}\tilde{L}_{k,l}^C\frac{f(|\boldsymbol{r}_i-\tilde{\boldsymbol{r}}_{k,l}|)}{\tilde{\rho}(\tilde{\boldsymbol{r}}_{k,l})},
\end{align}
where $\partial i$ is the set of grid points that are present within the cutoff distance $r_c$ from the particle $i$.
This operation is a simple redistribution of the Grad-CAM score depending on the contribution of each particles on the value of the density on each grid point $\tilde\rho_{k,l}$.

\subsection{Network architecture and hyperparameters}
\begin{figure}[htb]
    \centering
    \includegraphics[width=\textwidth]{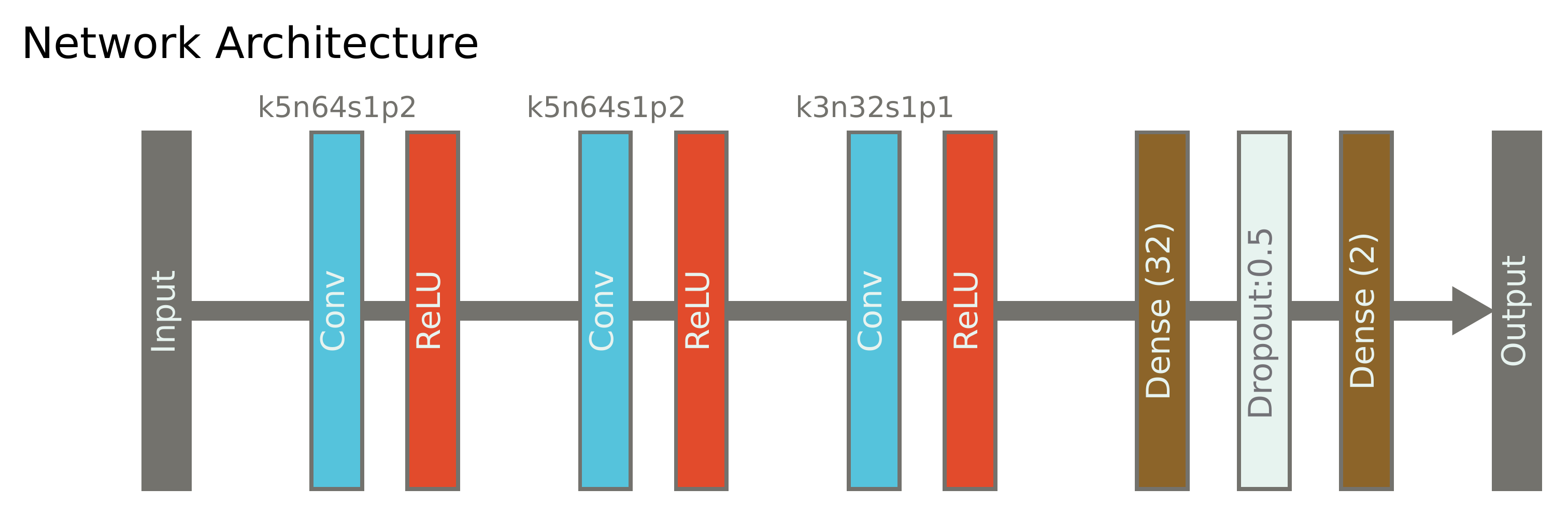}
    \caption{
    The sketch of the network architecture.
    The boxes with the letters Conv, ReLU, Dense and Dropout stand for the convolution layer, the ReLU activation, the fully connected layer and the dropour layer, respectively.
    For convolution layers, the kernel size, the number of filters, the stride, and the padding size are shown above the boxes (each value is shown after corresponding alphabets).
    For fully-connected layers, the number of outputs are shown in the parenthesis.
    For the dropout layer, the dropout ratio is set to be 0.5 as shown in the box.
    }
    \label{fig:network}
\end{figure}

Our network does not include any pooling layers and is simply composed of three convolution layers.
Each convolution layers are followed by the subsequent activations (the rectified linear units) and, after the third one, the fully-connected layer, the dropout layer, and the final fully-connected layer are stacked.
Note that although, at the learning stage, we apply the softmax function afterwards, the output layer of the network is the fully-connected layer in order to make it compatible with the Grad-CAM.
The schematic image of the network is given in Fig.~\ref{fig:network} where hyperparameters of each layer are also shown.

As for the convolution layers, we have employed the so-called circular-padding in order to take into account the periodic boundary conditions properly.
With the hyperparameters employed here, the feature maps obtained by each convolution layer all have the same size as that of the input data.
Therefore, we can obtain the fine-resolution Grad-CAM score field without any extra processing such as the guided-backpropagation.

\subsection{Learning protocols}
We employed the cross entropy as the loss function for the learning stage.
During the learning, we introduced the so-called $l2$-regularization to avoid the undesired overfitting.
These choices are made according to ref.~\cite{Swanson2020} where a classification problem between glasses and liquids has been tackled.
As for the optimization protocol, we used the Adam optimizer with the usual sets of parameters, $\beta_1=0.9$ and $\beta_2=0.999$.
Setting the initial learning rate as $l_r=10^{-6}$ and the batch size as $5$, we performed the learning with a fixed number of epochs: 250.
We reached these precise conditions after several trials and errors.
With these settings, the learning proceeded very smoothly and the network achieved almost 100\% accuracy without any symptom of the overfitting.

\subsection{Definition of angle-based indicators for the calculation of Tong-Tanaka order parameter}

\begin{figure}[htb]
    \centering
    \includegraphics[width=0.6\linewidth]{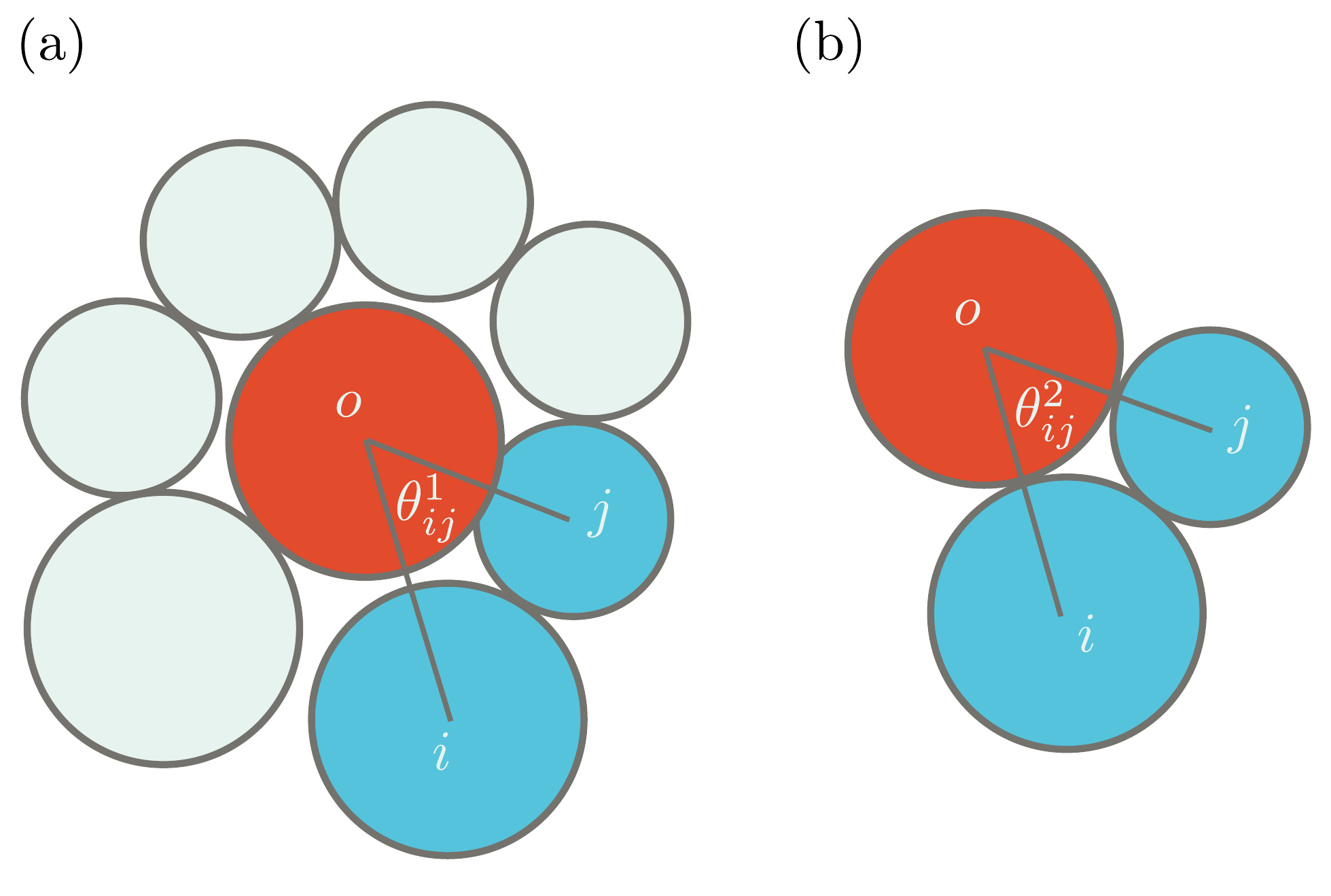}
    \caption{
    The sketches of the definitions of angles $\theta^1_{ij}$ and $\theta^2_{ij}$ that are used for the calculation of the TT-OP $\Theta_o$ for the particle $o$.
    (a) The one for $\theta^1_{ij}$ that is an angle realized by three particles ($o$, $i$, and $j$: particles $i$ and $j$ are the Voronoi neighbors of the particle $o$ and adjacent to each other as depicted) in the real configurations.
    (b) The one for $\theta^2_{ij}$ that is realized when all the particles are just in touch with each other (the distances between particles are all equal to the sum of radii for each pair).
    Note that, in this article, we employed the LJ parameter $\sigma_{ij}$ as "the sum of radii" for the particle pair $i$ and $j$.
    }
    \label{fig:TTOP}
\end{figure}

The bare Tong-Tanaka order parameter~\cite{TTNatCommun} for the particle $o$ is calculated using two angles $\theta^1_{ij}$ and $\theta^2_{ij}$ formed by the particle $o$ and its two neighboring neighbors $i$ and $j$ as:
\begin{align}
    \Theta_o=\frac{1}{N_o}\sum_{\langle ij\rangle}|\theta_{ij}^1-\theta_{ij}^2|,
\end{align}
where $\theta_{ij}^1$ is the angle realized in the samples by particles $i$, $j$ and $o$, and $\theta_{ij}^2$ is the ideal angle between the three particles that is realized when the distance between pairs are all equal to the interaction parameter $\sigma_{ij}$.
See Fig.~\ref{fig:TTOP} for the schematic pictures for these two angles.

\begin{figure}[tbh]
    \centering
    \includegraphics[width=\linewidth]{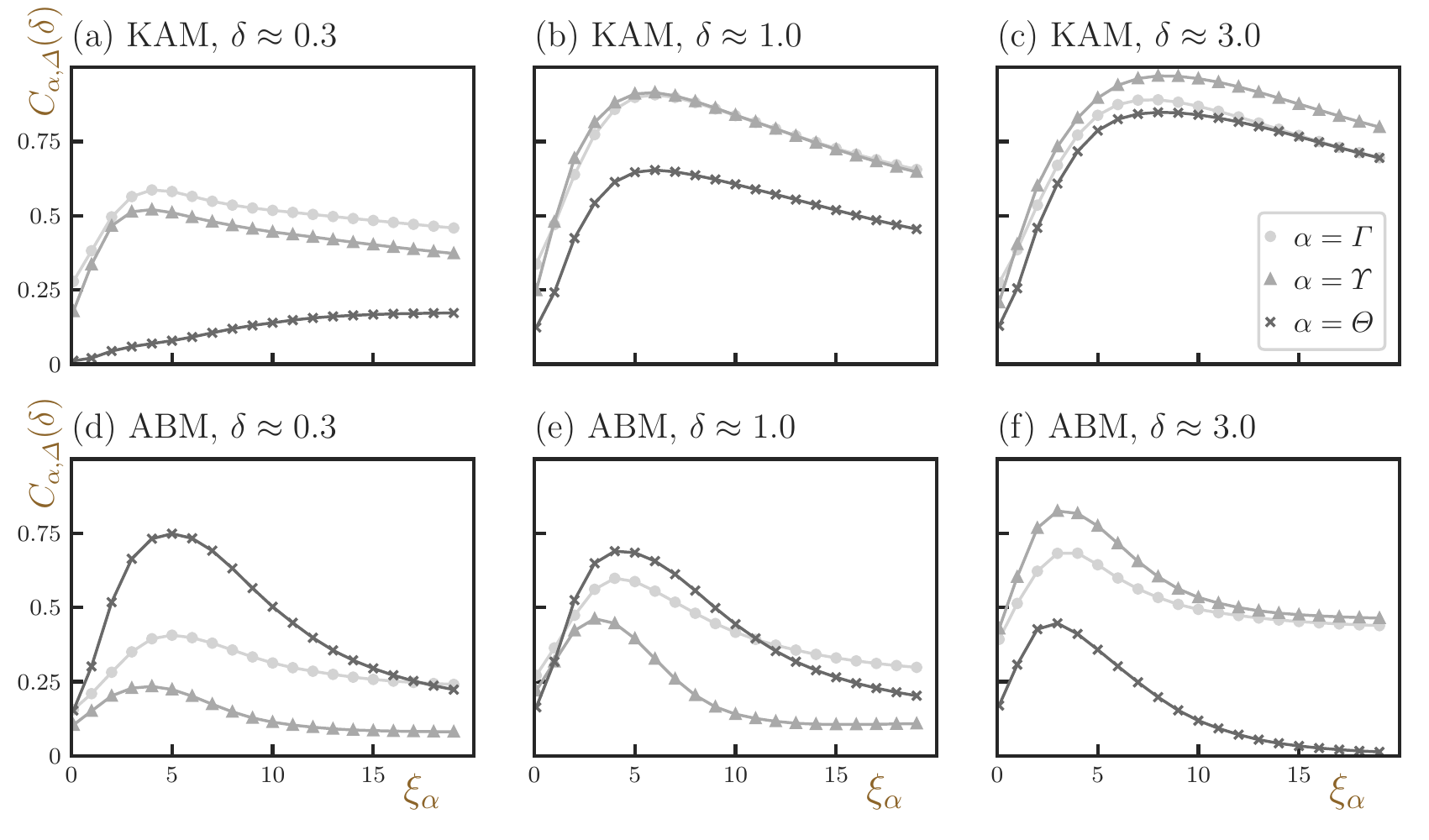}
    \caption{
    The dependence of the correlation coefficient between the structural indicators $\alpha(\alpha\in\lbrace\Gamma,\Upsilon,\Theta\rbrace)$ and the dynamic propensity $\Delta$ on the coarse-graining length for $\alpha$.
    As the dynamic propensity $\Delta$, the coarse-grained ones with the values of $\xi$ determined in Sec.~IV B are used.
    Different panels show results for different time scales and systems as shown in the panel titles, while different symbols distinguish different structural indicators as shown in the legend.
}
    \label{fig:xi}
\end{figure}

\subsection{Determination of coarse-graining lengths for structural indicators}
In this subsection, we investigate the dependence of the correlation coefficient between the dynamic propensity $\Delta$ and structural indicators $\alpha\in\lbrace\Gamma,\Upsilon,\Theta\rbrace$, $C_{\alpha,\Delta}$, on the coarse-graining length of the structural indicators $\alpha$, $\xi_\alpha$.
As the values of $\Delta$, we employed the coarse-grained ones with $\xi_\Delta$ determined in Sec.~IV B in the main text.
Thus, only the coarse-graining lengths $\xi_\alpha$ are varied.
In Fig.~\ref{fig:xi}, we present the results of three indicators ($\alpha\in\lbrace\Gamma,\Upsilon,\Theta\rbrace$) at three ``time'' scales ($\delta\approx 0.3, 1.0, 3.0$, which roughly correspond to $t\approx\tau_\alpha,3\tau_\alpha,10\tau_\alpha$) in concert with Figs.~1-3 in the main text.
The obtained optimal values $\xi^\ast_\alpha$ and the corresponding values of the correlation coefficients $C_{\alpha,\Delta}^{\rm max}$ and the time scale $\delta^\ast$ are summarized in Table~\ref{table:cmax}. 
Notice that, in this table \ref{table:cmax}, we present the optimal values among only three values of $\delta$ considered here ($\delta\approx 0.3,1.0,3.0$), unlike Table IV in the main text where the results of the investigation on the detailed $\delta$ dependence were presented.

As evident in Fig.~\ref{fig:xi} and Table~\ref{table:cmax}, the optimal values of $\xi_\alpha$ depend on both the indicators and time scales.
In the main text, for simplicity, we employed the largest value in Table~\ref{table:cmax} for each system regardless of the indicators $\alpha$ and the time scale $\delta$: that is, $\xi_\alpha=8.0$ for KAM and $\xi_\alpha=5.0$ for ABM.

We also note that $C_{\alpha,\Delta}$ at a longer-time scale does not always reach the maximum at a larger value of $\xi_\alpha$ than that for a shorter-time counterpart.
For example, in the case of the TT-OP in ABM, it shows the largest value of $C_{\Theta,\Delta}$ at a short time ($\delta\approx 0.3$) but the peak position ($\xi^\ast_\Theta=5.0$) is larger than the optimal $\xi_\Theta=3.0$ for a long-time dynamics ($\delta=3.0$).

\subsection{Results without coarse-graining of dynamic propensity}
In the main text, we coarse-grained the dynamic propensity $\Delta$ to smooth out the intra-domain noises in accordance with the coarse-grained structural indicators.
Such a coarse-graining operation of $\Delta$ results in higher values of the correlation coefficients $C_{\alpha,\Delta}$ compared to the case without coarse-graining.
In this subsection, we present the results without the coarse-graining of $\Delta$ as a reference.
We mention that in previous works, the dynamic propensity was not coarse-grained.

\begin{figure}[tbh]
    \centering
    \includegraphics[width=\linewidth]{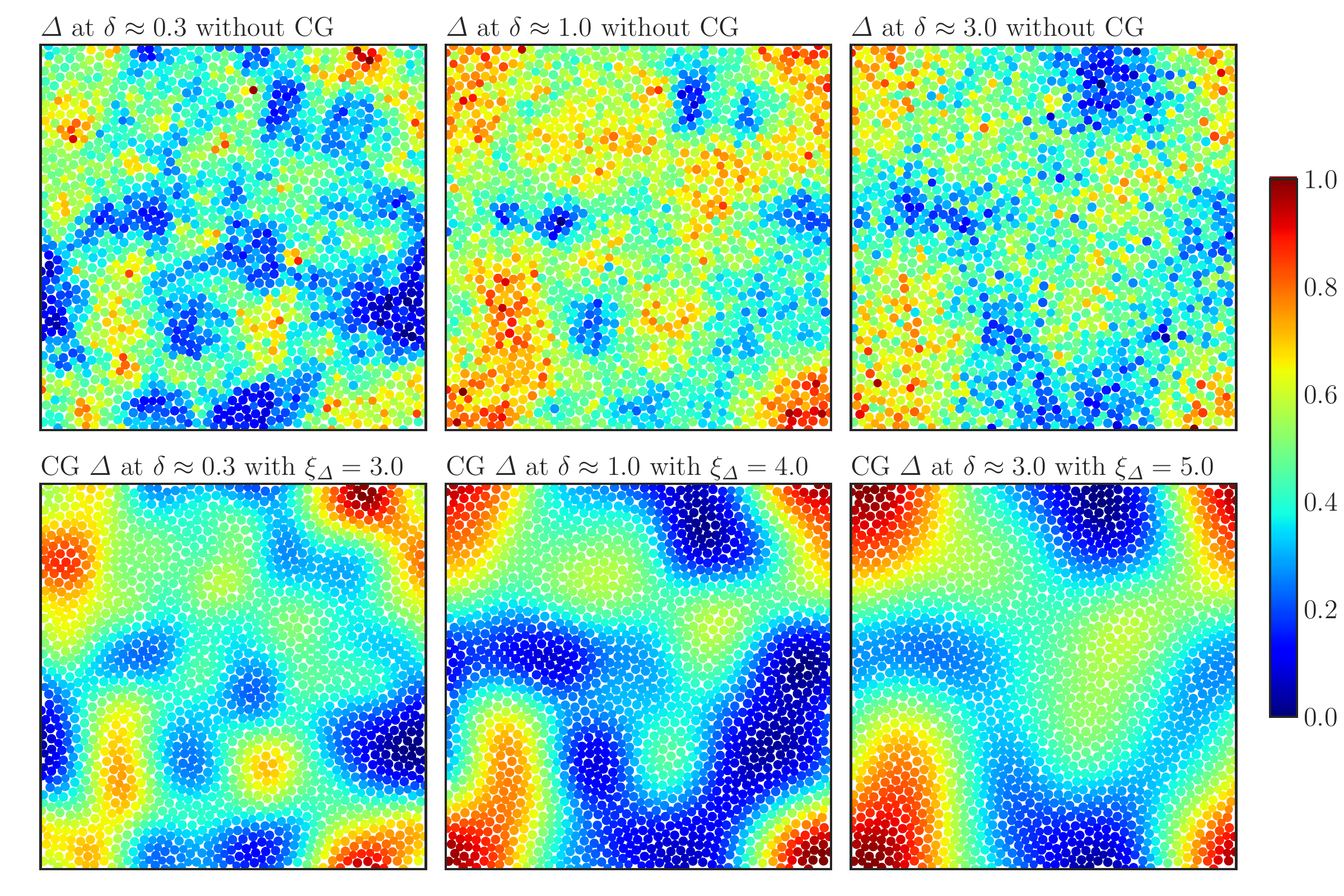}
    \caption{
Visualization of the dynamic propensity field at different time scales ($\delta\approx 0.3, 1.0, 3.0$ from left to right, as shown in the panel titles).
Results for the KAM are visualized.
The top row presents results without coarse-graining while the bottom row shows results with coarse-graining (the same ones as those presented in Fig.~1(d-f) in the main text).
}
    \label{fig:prop_KAM}
\end{figure}

\begin{figure}[tbh]
    \centering
    \includegraphics[width=\linewidth]{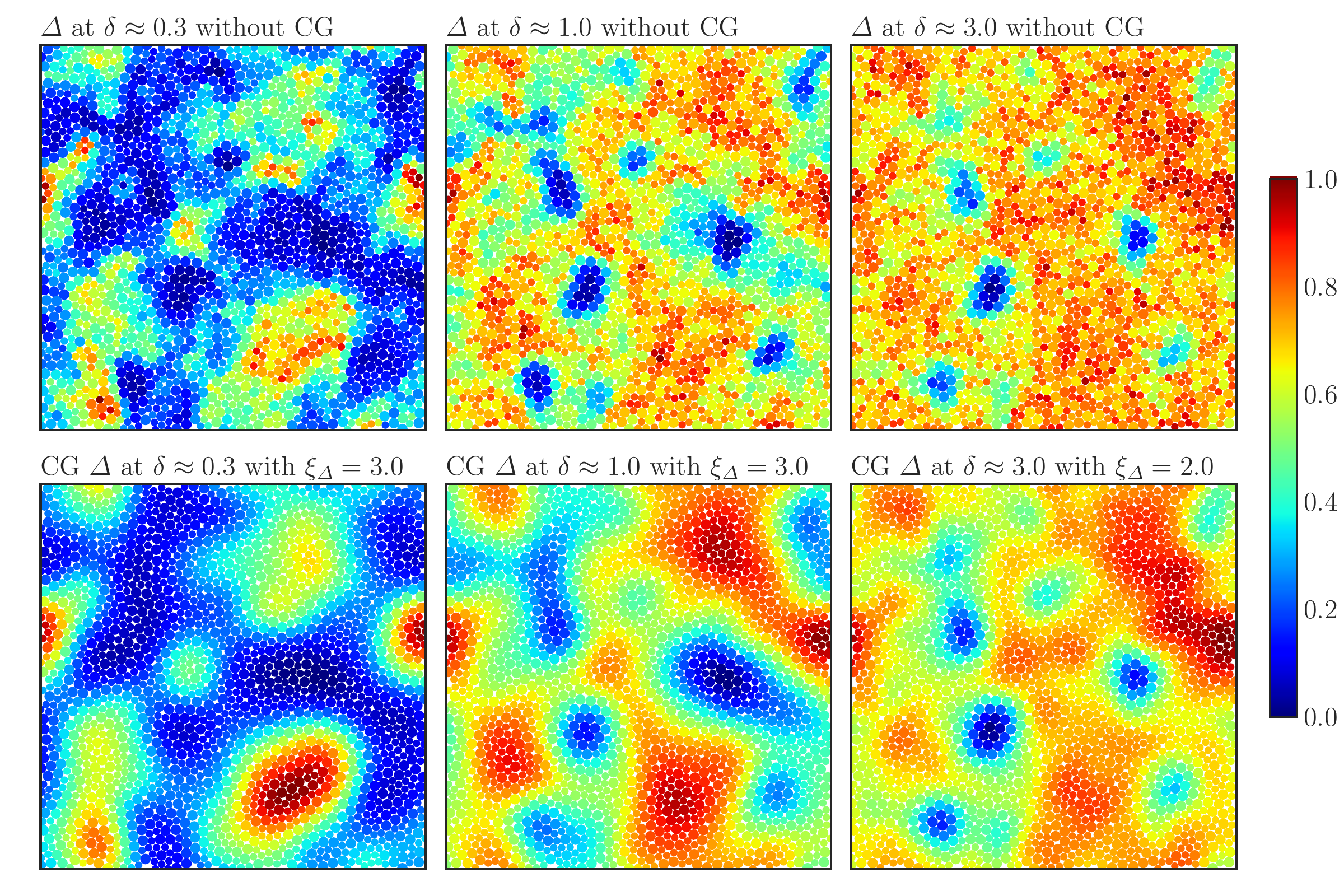}
    \caption{
    The same information as that in Fig.~\ref{fig:prop_KAM} but for the ABM.
    The bottom row corresponds to Fig.~\ref{fig:prop_ABM}(d-f) in the main text.
}
    \label{fig:prop_ABM}
\end{figure}

In the top row of Fig.~\ref{fig:prop_KAM}, we show the visualization of the bare (non-coarse-grained) dynamic propensity field $\Delta$ in the KAM.
For ease of comparison, we also present the coarse-grained $\Delta$ field (the same ones presented in Fig.~1 in the main text) in the bottom panels.
From this figure, we can tell that the coarse-graining with our choices of $\xi_\Delta$ merely smooths out the noises inside the mobile(red)/immobile(blue) domains and the mesoscopic spatial patterns remain unchanged.
We show the same information for the ABM in Fig.~\ref{fig:prop_ABM}.

\begin{figure}[tbh]
    \centering
    \includegraphics[width=\linewidth]{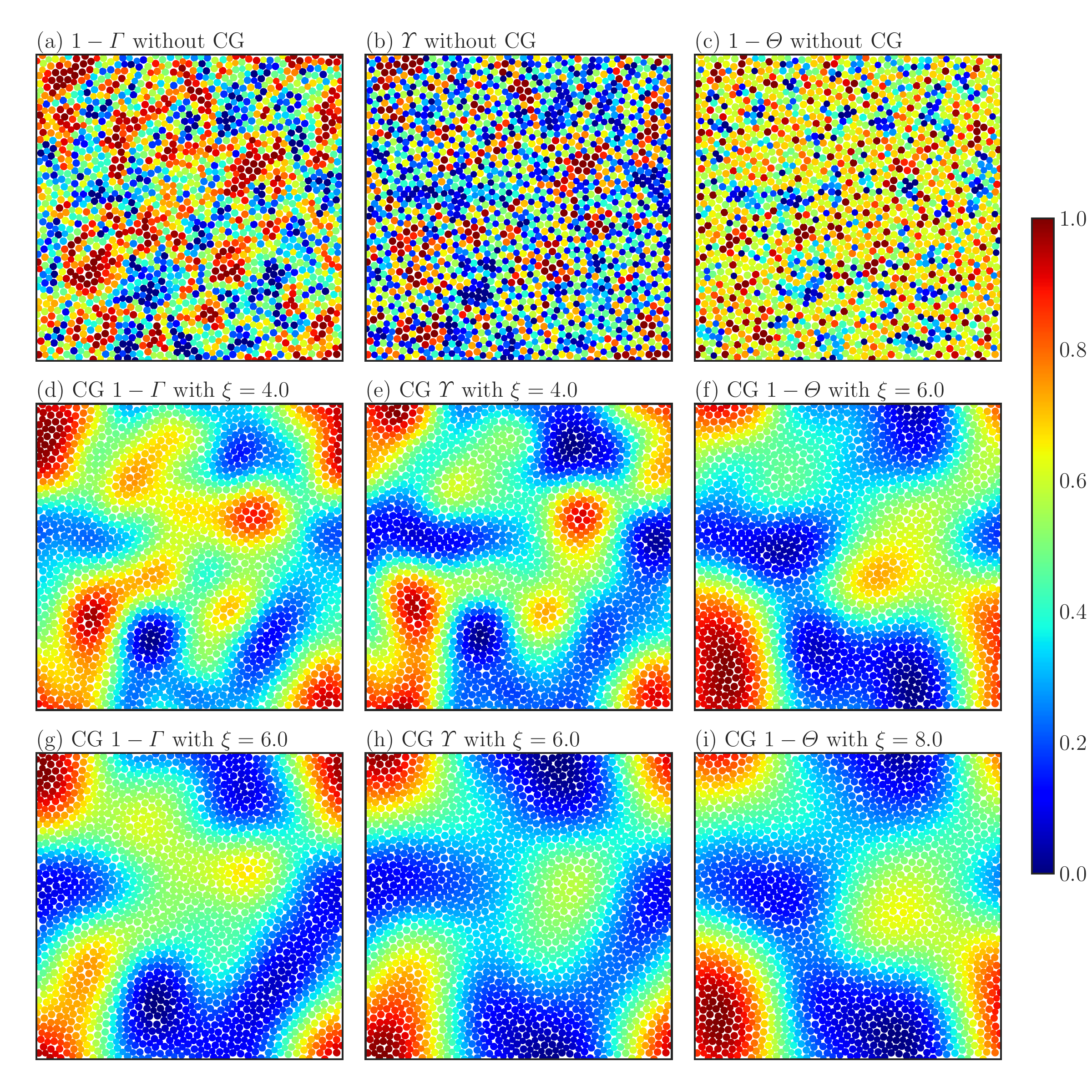}
    \caption{
Visualization of three different structural indicators  ($1-\Gamma, \Upsilon, 1-\Theta$ from left to right, as shown in the panel titles).
Results for the KAM are visualized.
Top, middle, and bottom rows present results without coarse-graining, those with $\xi$ that maximize $C_{\alpha,\Delta}$ with the non-coarse-grained $\Delta$ and those with $\xi$ that maximize $C_{\alpha, \Delta}$ with the coarse-grained $\Delta$, respectively.}
    \label{fig:str_KAM}
\end{figure}

\begin{figure}[tbh]
    \centering
    \includegraphics[width=\linewidth]{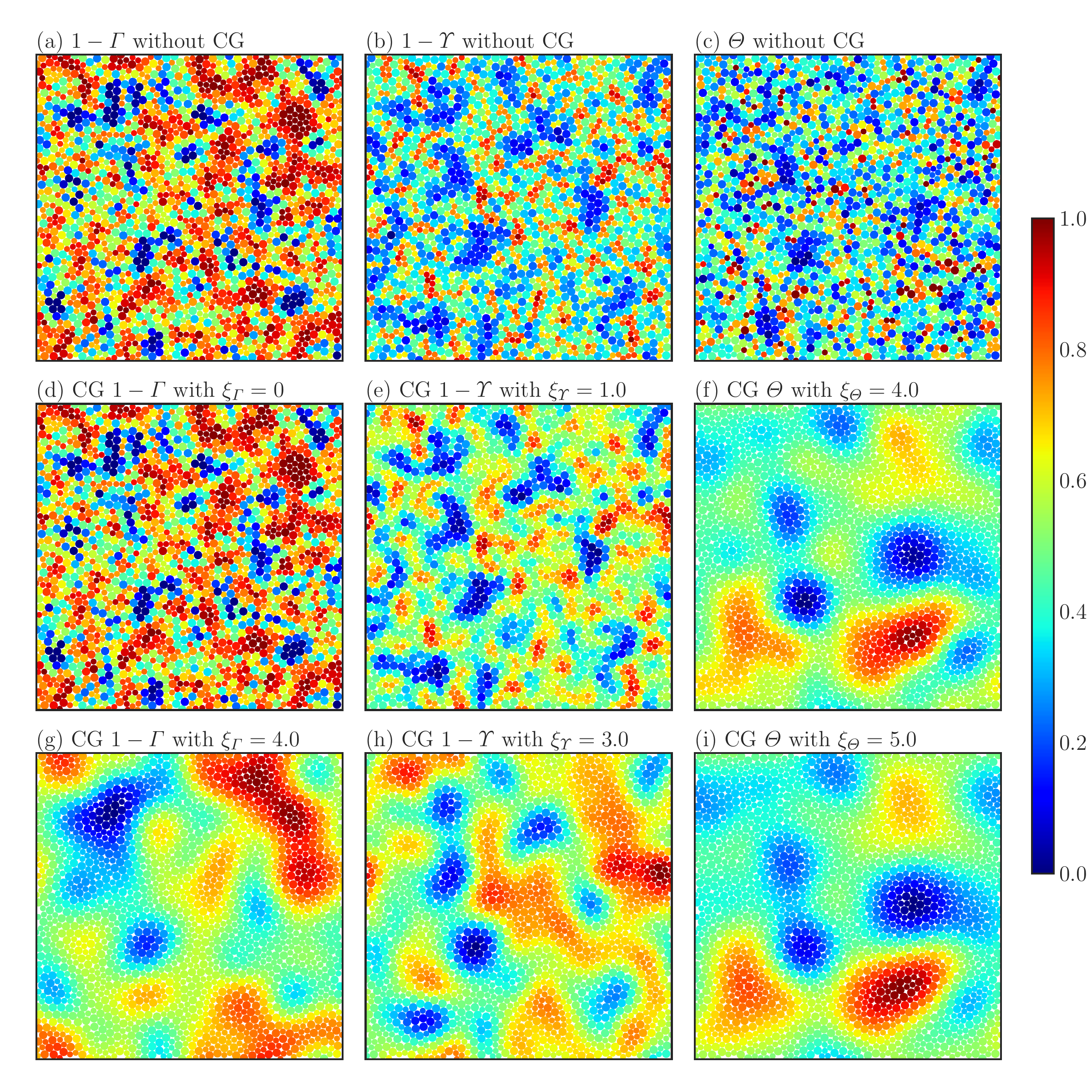}
    \caption{
The same information as that in Fig.~\ref{fig:str_KAM} but for the ABM.
Also, instead of $\Upsilon$ and $1-\Theta$, $1-\Upsilon$ and $\Theta$ are visualized.}
    \label{fig:str_ABM}
\end{figure}

We further present the visualization of the structural indicators without coarse-graining in Fig.~\ref{fig:str_KAM} and \ref{fig:str_ABM} for KAM and ABM respectively.
The upper rows present the bare structural indicators and the center and bottom rows show coarse-grained fields.
In particular, for center row, the coarse-graining lengths that maximize the correlation with the bare propensity $\Delta$ are used.
For the bottom rows, $\xi_\alpha^\ast$ that maximizes the correlation with the coarse-grained $\Delta$ are employed (still, different optimal $\xi_\alpha^\ast$ are used for different indicators, and thus different information from the Figs.~1 and 2 in the main text is presented).
Note that, for ease of visibility, and for the visualization of indicators without coarse-graining, we normalized the values excluding extreme values.

\begin{figure}[tbh]
    \centering
    \includegraphics[width=\linewidth]{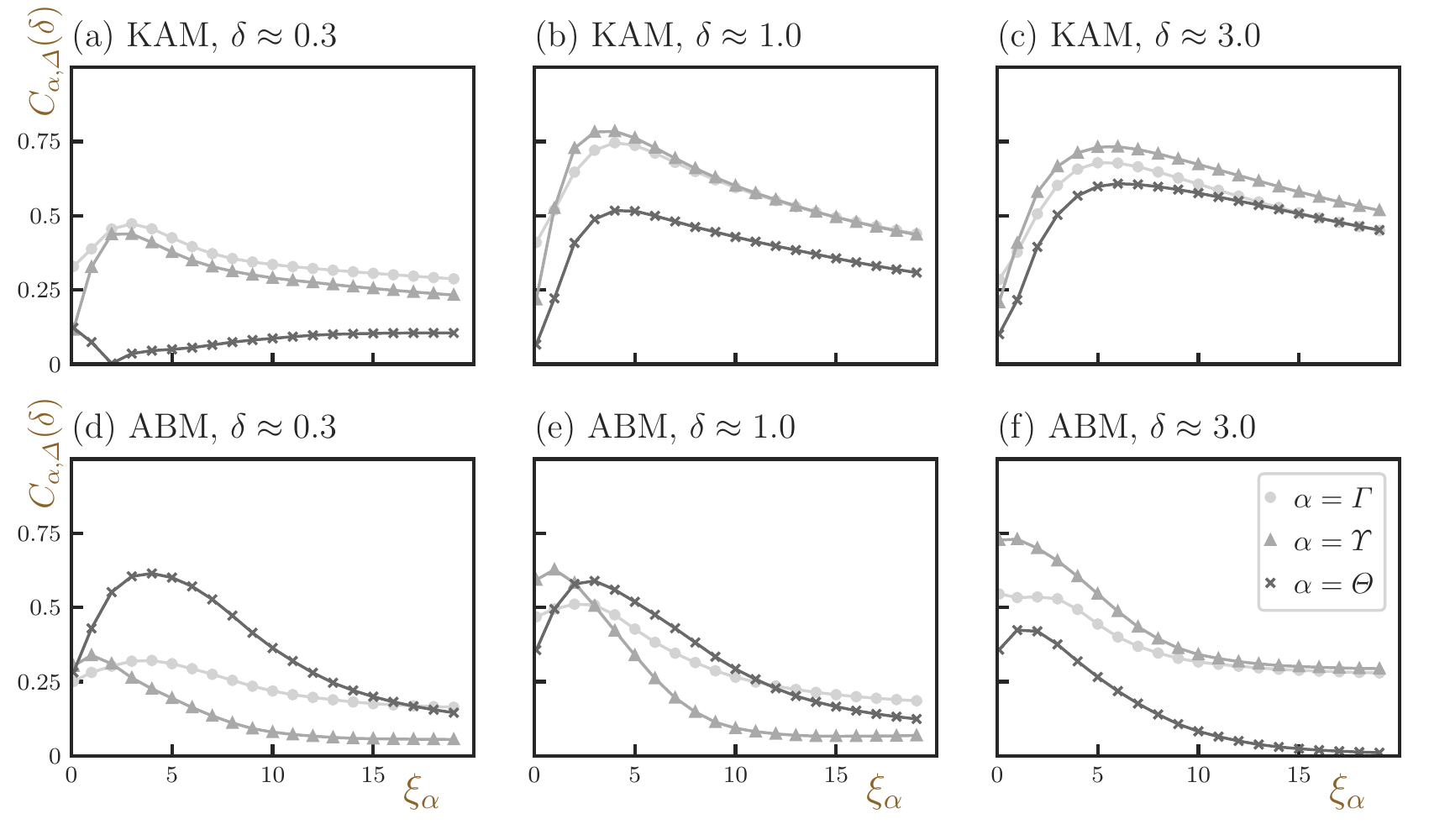}
    \caption{
The same information as that in Fig.~\ref{fig:xi} but with the non-coarse-grained dynamic propensity.
}
    \label{fig:xi_WOCG}
\end{figure}

In Fig.~\ref{fig:xi_WOCG}, we further plot the $\xi_\alpha$ dependence of $C_{\alpha, \Delta}$ as those in Fig.~\ref{fig:xi}.
Just one difference is that the bare propensity fields are employed as $\Delta$ in Fig.~\ref{fig:xi_WOCG}.
Compared to the results with coarse-grained $\Delta$ presented in Fig.~\ref{fig:xi}, the values of $C_{\alpha,\Delta}$ are smaller as a whole.
Also, not only the peak heights but also the peak positions differ from the results in Fig.~\ref{fig:xi}.
We summarized the results of Fig.~\ref{fig:xi_WOCG} in Table~\ref{table:cmax} as well.

\tabcolsep = 8pt
\begingroup
\renewcommand{\arraystretch}{1.5}
\begin{table*}[tbh]
  \caption{Maximum values of the correlation coefficient and optimal values of $\delta$ and $\xi$}
  \centering\label{table:cmax}
  \begin{tabular}{c|ccc|ccc|ccc|ccc}
 
    \hline
     &\multicolumn{6}{c|}{with CG of $\Delta$}  &\multicolumn{6}{c}{without CG of $\Delta$} \\
    \cline{2-13} 
     &\multicolumn{3}{c|}{KAM} &\multicolumn{3}{c|}{ABM}  &\multicolumn{3}{c|}{KAM} &\multicolumn{3}{c}{ABM} \\
     \cline{2-13}
     &$C_{\alpha,\Delta}^{\rm max}$ &$\delta^\ast_\alpha$ &$\xi^\ast_\alpha$     &$C_{\alpha,\Delta}^{\rm max}$ &$\delta^\ast_\alpha$ &$\xi^\ast_\alpha$     &$C_{\alpha,\Delta}^{\rm max}$ &$\delta^\ast_\alpha$ &$\xi^\ast_\alpha$     &$C_{\alpha,\Delta}^{\rm max}$ &$\delta^\ast_\alpha$ &$\xi^\ast_\alpha$ \\
     \hline 
    Grad-CAM score $\Gamma$ &0.907 &1.0&6.0 &0.683 &3.0 & 4.0&0.746 &1.0&4.0&0.546&3.0&0\\
    Voronoi volume $\Upsilon$ &0.971 &3.0&8.0 &0.825 &3.0 & 3.0&0.784 &1.0&4.0&0.730&3.0&1.0\\
    TT-OP $\Theta$ &0.848 &3.0&8.0 &0.749 &0.3 & 5.0&0.608 &3.0&6.0&0.615&0.3&4.0\\
    \hline
  \end{tabular}
\end{table*}
\endgroup
\clearpage
\section{Fundamental information about the samples}
\subsection{Evolution of potential energy during cooling}\label{sec:energy}
To evidence that the crystallization has been avoided during the cooling process for the sample preparation, we show the time evolution of the potential energy $E$ as a function of the temperature for both the ABM and the KAM in Figs.~\ref{fig:energy}(a) and (d) respectively.
From these figures, we can tell that no discontinuous jumps occur in the temperature-energy curves in both systems under the given cooling rate and the amorphous samples at a very low temperature $T=0.05$ are obtained as we intended.

\subsection{Potential energy of the corresponding inherent structures}
In Figs.~\ref{fig:energy}(b) and (d), we plot the potential energy $E_0$ of the so-called inherent structures that can be obtained by minimizing the potential energy of instantaneous configurations.
In these plots, the inherent structure energy $E_0$ is again plotted as a function of the temperature of the original instantaneous configurations.
Sastry and coworkers~\cite{Sastry1998} have reported that, when cooling, this $E_0$ starts decreasing as the glass-like slowing-down sets in.
In this article, we chose the temperature at which the $E_0$ starts decreasing as the one for the liquid samples (in the case of the KAM, at odds with the three-dimensional system that Sastry investigated~\cite{Sastry1998}, $E_0$ decreases gradually even at high temperatures in 2D systems~\cite{Swanson2020}: we chose the temperature at which the decreasing tendency becomes almost linear: see Fig.~\ref{fig:energy}).

\subsection{Empirical definition of the glass transition point}
\begin{figure}[h]
    \centering
    \includegraphics[width=\linewidth]{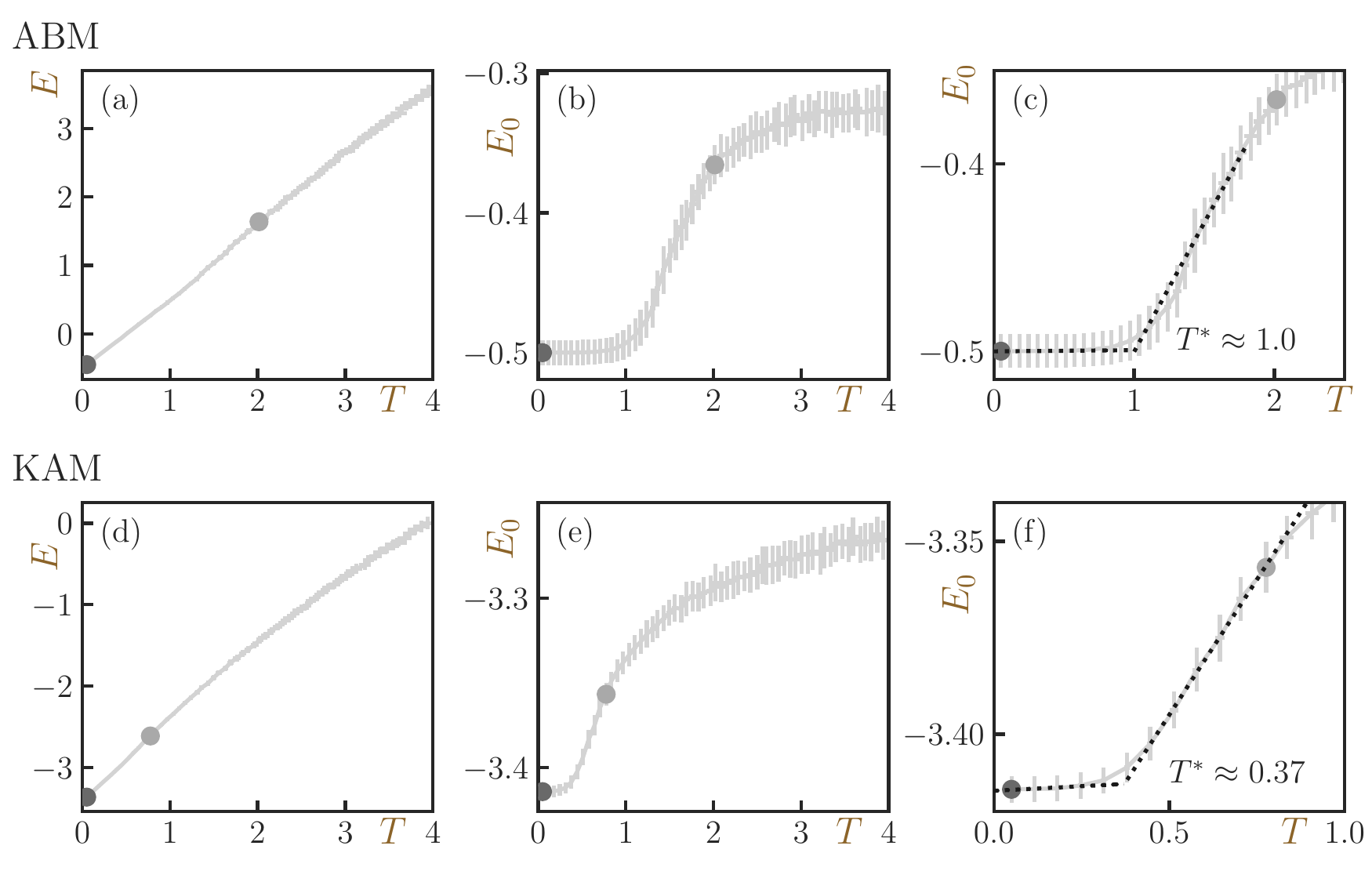}
    \caption{
    The evolution of the potential energy as functions of the temperature $T$ for the ABM (a-c) and the KAM (d-f) systems.
    (a,d) The instantaneous potential energy $E$.
    (b,e) The potential energy of the inherent structure $E_0$.
    (c,f) The close-up plots of $E_0$ around the glass transition point $T^\ast$.
    The dark and light-gray circles depict the values of the $T_{\rm G}$ and $T_{\rm L}$, respectively (see the main text for the definitions of these temperatures).
    The dotted lines in panels (c,f) are the linear fits for the low and intermediate temperature regimes.
}
    \label{fig:energy}
\end{figure}

We further define the empirical glass transition point under our cooling rate ($\dot{T}\approx 8.33\times 10^{-5}$) relying on the qualitative change in the temperature dependence of $E_0$: we fitted the low-temperature regime where $E_0$ becomes almost constant down to $T=T_{\rm G}=0.05$ and the middle-temperature regime where $E_0$ changes linearly as a function of $T$ and the intersection of these two lines is defined as the glass transition point $T^\ast$.
The obtained values are $T^\ast\approx 1.0$ for the ABM and $T^\ast\approx 0.37$ for the KAM.
See Figs.~\ref{fig:energy}(c) and (f) for the close-up plots around the glass transition point $T^\ast$.

\subsection{Radial distribution function}
\begin{figure}[h]
    \centering
    \includegraphics[width=0.6\linewidth]{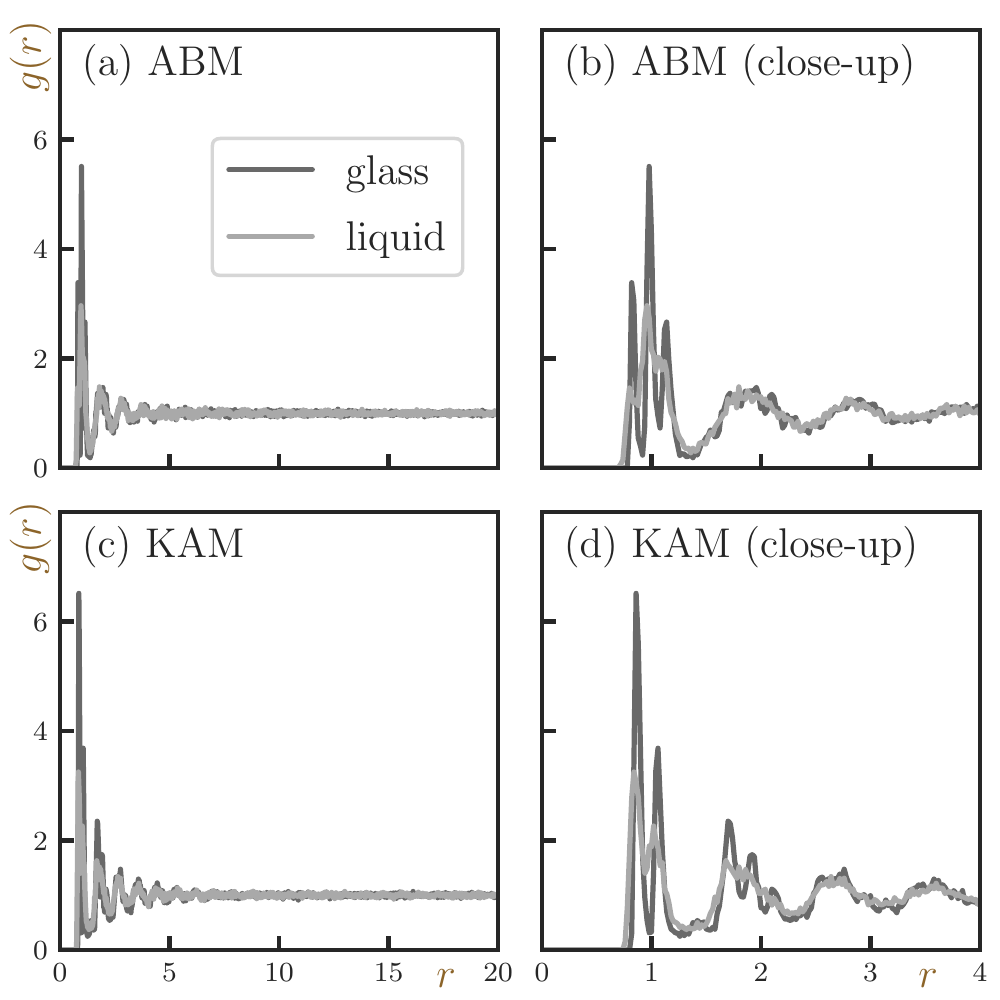}
    \caption{
    The radial distribution functions $g(r)$ for the ABM (a, b)  and the KAM (c,d) systems.
    Panels (b) and (d) present the close-up plots to show the development of the short-range local structure in glass samples.
    Different colors of lines distinguish glasses and liquids samples as shown in the legend.
    }
    \label{fig:RDF}
\end{figure}

In Sec.~\ref{sec:energy} in this supplemental material, we showed that judging from the evolution of the total potential energy $E$, the system seems not to experience crystallization at least at the global level.
In this subsection, we also show that the crystallization is indeed avoided by directly measuring the radial distribution function (RDF) $g(r)$.
We present the RDFs for both the ABM and KAM systems in Fig.~\ref{fig:RDF} for both the liquid ($T=T_L$) and the glass ($T=T_G$) configurations.
Although the glass systems exhibit more developed local structures compared to the liquid ones at short-range distances, we do not see any sign of global crystallization.

\newpage
\section{Average and standard deviation of correlation coefficients}
The averages and the standard deviations of the correlation functions of different structural indicators are summarized in Table~\ref{table:ave_and_std_corre}.
In particular, in this table, in addition to the values for the Pearson's coefficients that are discussed in the main text, we present the results for Spearman's coefficients in parentheses.
In terms of these statistical variables, the Pearson's and the Spearman's definitions give the semi-quantitatively same results.

\tabcolsep = 5pt
\begingroup
\renewcommand{\arraystretch}{1.5}
\begin{table*}[tbh]
  \caption{Correlation coefficients between different structural indicators}
  \centering\label{table:ave_and_std_corre}
  \begin{tabular}{c|ccc}
 
    \hline
     & $C_{\Gamma,\Upsilon}$ (GC-S vs. VV) & $C_{\Gamma,\Theta}$ (GC-S vs. TT-OP)  & $C_{\Theta,\Upsilon}$ (TT-OP vs. VV) \\
    \hline 
    KAM (G) &$-0.828\pm 0.055$ ($-0.816\pm 0.061$) &$0.426\pm 0.143$ ($0.407\pm 0.143$) &$-0.597\pm 0.111$ ($-0.580\pm 0.116$)\\
    KAM (L) &$0.857\pm 0.047$ ($0.847\pm 0.052$) &$-0.014\pm 0.179$ ($-0.016\pm 0.177$) &$-0.201\pm 0.170$ ($-0.192\pm 0.169$)\\
    ABM (G) &$0.585\pm 0.094$ ($0.539\pm 0.093$) &$-0.378\pm 0.121$ ($-0.311\pm 0.111$) &$-0.327\pm 0.129$ ($-0.256\pm 0.116$)\\
    ABM (L) &$0.036\pm 0.180$ ($0.038\pm 0.178$) &$0.292\pm 0.110$ ($0.278\pm 0.109$) &$-0.225\pm 0.125$ ($-0.196\pm 0.118$)\\
    \hline
  \end{tabular}
\end{table*}
\endgroup

\newpage
\bibliography{supplement}